\documentclass{aa}    
\titlerunning{Mg and Si}
\bibpunct{(}{)}{;}{a}{}{,}
\usepackage{natbib}
\usepackage[colorlinks=true, citecolor=blue]{hyperref}
\usepackage{xcolor}
\usepackage{longtable}
\usepackage{graphicx}
\usepackage{txfonts}
\begin{document}

\title{The CARMENES search for exoplanets around M dwarfs\fnmsep\thanks{Table~\ref{tab:sample} is only available in electronic form at the CDS via anonymous ftp to cdsarc.u-strasbg.fr (130.79.128.5) or via http://cdsweb.u-strasbg.fr/cgi-bin/qcat?J/A+A/.}}

\subtitle{Magnesium and silicon abundances of K7-M5.5 stars}

   \author{H.~M.~Tabernero
          \inst{1,2}
          \and
           Y.~Shan\inst{3,4}
          \and
          J.~A.~Caballero\inst{5}
          \and
         C.~Duque-Arribas\inst{1}
          \and          
          D.~Montes\inst{1}
          \and
          J.~I.~Gonz{\'a}lez Hern{\'a}ndez\inst{6,7}
          \and
          M.~R.~Zapatero Osorio\inst{5}
          \and
          A.~Schweitzer\inst{8}
          \and
          Th.~Henning\inst{9}
          \and
          M.~Cortés-Contreras\inst{1}
          \and
          A.~Quirrenbach\inst{10}
          \and
          P.~J.~Amado\inst{11}
          \and
          A.~Reiners\inst{3}
          \and
          I.~Ribas\inst{12,13}
          \and 
           G.~Bergond\inst{14}
           \and
          J.~C. Morales\inst{12,13} 
          }
          
\institute{
  Departamento de F{\'i}sica de la Tierra y Astrof{\'i}sica \& 
           IPARCOS-UCM (Instituto de F\'{i}sica de Part\'{i}culas y del Cosmos de la UCM), 
           Facultad de Ciencias F{\'i}sicas, Universidad Complutense de Madrid, 28040 Madrid, Spain \\
     \email{htabernero@ucm.es}
\and        
Centro de Astrobiología, CSIC-INTA, Carretera de Ajalvir km 4, 28850 Torrejón de Ardoz, Madrid, Spain
\and
           Institut f\"{u}r Astrophysik und Geophysik, Georg-August-Universit\"{a}t-G\"{o}ttingen, 
           Friedrich-Hund-Platz 1, 37077 G\"{o}ttingen, Germany 
           \and
             Centre for Planetary Habitability, Department of Geosciences, University of Oslo, Sem Saelands vei 2b 0315 Oslo, Norway
           \and
           Centro de Astrobiología, CSIC-INTA, Camino Bajo del Castillo s/n, 28691 Villanueva de la Cañada, Madrid, Spain
           \and
           Universidad de La Laguna, Departamento de Astrof\'{i}sica,
           38206 La Laguna, Tenerife, Spain
           \and
           Instituto de Astrof\'{i}sica de Canarias,
           c/ V\'{i}a L\'{a}ctea s/n, 38205 La Laguna, Tenerife, Spain
        \and
           Hamburger Sternwarte,
           Gojenbergsweg 112, 21029 Hamburg, Germany
           \and 
              Max-Planck-Institut f\"{u}r Astronomie, K\"{o}nigstuhl 17, 69117 Heidelberg, Germany
           \and
           Landessternwarte, Zentrum f\"{u}r Astronomie der Universit\"{a}t Heidelberg,
           K\"{o}nigstuhl 12, 69117 Heidelberg, Germany
           \and
           Instituto de Astrof\'{i}sica de Andaluc\'{i}a (IAA-CSIC), 
           Glorieta de la Astronom\'{i}a s/n, 18008 Granada, Spain          
              \and
           Institut de Ci\`{e}ncies de l'Espai (CSIC), Campus UAB, c/ de Can Magrans s/n, 
           08193 Cerdanyola del Vall\`{e}s, Spain
           \and
           Institut d'Estudis Espacials de Catalunya (IEEC), c/ Gran Capit\`{a} 2-4, 
           08034 Barcelona, Spain
           \and
           Centro Astron{\'o}mico Hispano-Alem{\'a}n, Observatorio de Calar Alto, Sierra de los Filabres, 04550 G\'{e}rgal, Almer\'{i}a, Spain
           }
   
\date{Received 20 03 2024 / Accepted 15 07 2024}

% \abstract{}{}{}{}{}
% 5 {} token are mandatory
 
  \abstract
  %    context heading (optional)

  \abstract{We present the abundances of magnesium (Mg) and silicon (Si) for  314 dwarf stars with spectral types in the interval  K7.0--M5.5  ($T_{\rm eff}$ range $\approx$ 4200--3050~K) observed with the CARMENES high-resolution spectrograph at the 3.5 m telescope at the Calar Alto Observatory. Our analysis employs the BT-Settl model atmospheres, the radiative transfer code {\tt Turbospectrum}, and a state-of-the-art selection of atomic and molecular data. These Mg and Si abundances are critical for understanding both the chemical evolution and assembly of the Milky Way and the formation and composition of rocky planets.  Our chemical abundances show a line-to-line scatter at the level of 0.1\,dex for all studied spectral types. The typical error bar of our chemical abundance measurements is $\pm$~0.11~dex (Mg) and $\pm$~0.16~dex (Si) for all spectral types based on the comparison of the results obtained for stellar components of multiple systems. The derived abundances are consistent with the galactic evolution trends and observed chemical abundance distribution of earlier FGK-type stars  in the solar neighbourhood. Besides, our analysis provides compatible abundances for stars in multiple systems. In addition, we studied the abundances of different galactic stellar populations. In this paper, we also explore the relation of the Mg and Si abundances of stars with and without known planets.} 

   \maketitle
%
%________________________________________________________________

\section{Introduction}

The spectra of late K and M dwarf stars pose a challenge to modern stellar astrophysics. Despite their complex spectra, the precise determination of chemical abundances by means of high-resolution spectroscopy is critical to our understanding of the formation and assembly of the Milky Way \citep[e.g.][]{ben11,mit14,sil15,hog16,bud21,gil22,ran22}, as well as the chemical composition and formation of exoplanets \citep[see ][and references therein]{gon10,gon13,adi12,brew16,adi21}. Among the chemical elements, magnesium  (Mg) and silicon (Si) stand out as hallmarks of stellar and planetary science. These elements are mostly produced in massive stars before being ejected into the interstellar medium by core-collapse supernovae through the $\alpha$-process \citep[SNe II, Ib, and Ic;][]{kob20}. Furthermore, Mg and Si belong to the family of the  refractory elements, alongside Fe, and are believed to be the building blocks of rocky planets \citep{morand80}. Due to this fact, the abundances of Mg and Si are a key to model the structure of rocky planets as shown in previous works like, for example: \citet{dorn17} and \citet{lich21}. To this aim, the surface abundances of the host star can be used as a proxy to infer the properties of the rocky planets orbiting them as shown by \citet{del10}, \citet{adi15}, \citet{brew16b}, \citet{san17}, \citet{sua18}, \citet{adi21}, \citet{cab22}, among others. During the last two decades the study of Mg and Si abundances using high-resolution spectra has targeted mostly FGK stars \citep[see e.g.][]{ben03,adi12,wei19}, and is now moving towards cooler targets like the M-dwarf stars. All in all, the M-dwarf stars are interesting targets for abundance characterisation since they represent about 70--75\% of the stellar population of our Galaxy \citep{hen06,win15,rey21}. They also have main-sequence lifetimes exceeding the current known age of the Universe, due to the slow fusion processes that take place in their (mainly) convective interiors \citep{ada97}.\\

These stars have effective temperatures ($T_{\rm eff}$) in the range 2\,300--4\,000~K \citep{kir91,cif20} that give rise to several molecules in their atmospheres. The spectral features due to these molecules dominate their optical and infrared spectra and mask the stellar pseudo-continuum, thus making the abundance determination more challenging than for FGK stars. In recent years, some abundance studies of chemical elements other than iron are moving in the direction of the analysis of cooler M dwarf stars. For instance, \citet{mon18} computed the abundances of 13 chemical elements of 192 FGK stars with an M-dwarf companion. In addition, \citet{mal20} determined abundances for a number of elements, including Mg and Si, for 204 M dwarfs. They did not study individual line profiles, but instead employed a Bayesian and principal component analysis based on a calibration using FGK stars with an M-star companion.\\

The near-infrared (NIR) represents the peak of the M dwarf spectral energy distribution and the now-preferred wavelength range for chemical abundance studies in these stars. Several recent works have shown that the spectral synthesis method on high-resolution NIR spectra is effective for chemical analysis of cool dwarfs. A handful of M dwarfs have Mg and Si measured in this manner. \citet{ish20} determined the chemical abundances of eight elements (including Mg) for five M dwarfs using NIR spectra (9\,600--17\,100~\AA{}) acquired with the CARMENES spectrograph \citep{quir20}. $H$-band spectra ($\sim$~15\,000--17\,000~\AA{}) from the SDSS-APOGEE survey \citep{maj17} was used to characterise a palette of chemical elements, covering Mg and Si, for up to 24 M dwarfs \citep{sou17,sou18,sou22}. Likewise, \citet{ish22} provided Mg abundances for 13 M dwarfs and Si for one M dwarf using Subaru/IRD spectra (9\,700--17\,500~\AA{}). More recently, \citet{jah23} used SPIRou spectra (9\,800--23\,500~\AA{}) for a comprehensive chemical study of Barnard's Star.  while, \citet{hej23} used the IGRINS spectrograph (14\,500--24\,500~\AA{}) to characterise WASP-107.  More However, one key challenge with scaling up these studies has been the dearth of high-quality data.\\

The Calar Alto high-Resolution search for M dwarfs with Exoearths with Near-infrared and optical Échelle Spectrographs \citep[CARMENES\footnote{\url{https://carmenes.caha.es/}}, see][]{quir20} has been monitoring more than 300 M dwarfs during the last years. CARMENES is a dual-channel spectrograph mounted at the 3.5~m telescope at Calar Alto Observatory (Almer{\'i}a, Spain). CARMENES represents a step forward to study the planet occurrence and other science cases such as the computation of chemical abundances of the M dwarfs because of its high spectral resolution and its wide wavelength coverage \citep{quir14}. In short, the CARMENES survey has provided high-quality spectra for hundreds of nearby M dwarfs \citep{rib23,nag23}. Some of these spectroscopic data have been used to study other elemental abundances such as those of vanadium, rubidium, strontium, and zirconium \citep{abi20,sha21}. In this work, we present the study of the magnesium and silicon abundances for 314 stars observed by the CARMENES consortium. We show that six \ion{Mg}{i} and six \ion{Si}{i} atomic lines that fall in the CARMENES wavelength range can be reproduced with the latest available atomic and molecular parameters and state-of-the-art stellar atmospheric models. Finally, the homogeneously derived Mg and Si abundances of the late-K and M dwarfs expand the results shown in the pilot study of \citet{ish20}. The newly derived abundances are compatible with the abundance patterns seen for earlier spectral types \citep[see e.g.][]{adi12}. This paper is structured as follows: the reduction of the CARMENES data and the computation of the Mg and Si abundances are presented in Sect.~\ref{sec:dataproc}. The results and subsequent discussion on the determined abundances are given in Sect.~\ref{sec:abundances}, while their implications for exoplanet science are given in Sect.~\ref{sec:imp}. Finally, the summary and conclusions of this work are given in Sect.~\ref{sec:conclusions}. 

\section{Data processing}
\label{sec:dataproc}
\subsection{Stellar sample}
\label{sec:obs}

Our sample contains 314 nearby K and M dwarfs listed in Table~\ref{tab:sample}. These stars were observed within the framework of the CARMENES guaranteed time observations \citep[GTO, see][]{rei18,rib23} and legacy programs. Their spectra were reduced according to the standard CARMENES GTO data flow with the {\tt caracal} and {\tt serval} pipelines, which compute a single co-added spectrum per observation and instrument channel \citep{Cab16b,Zech18}. These individual spectra were corrected from telluric absorption using {\tt molecfit} \citep{sme15,kau15} and combined into a single high S/N template spectrum for each individual star in our sample. Further details on the telluric correction and how these template spectra were computed can be found in \citet{nag23}. After the telluric correction, we applied the vacuum-to-air conversion given by \citet{mor00}, followed by the Doppler correction into the laboratory rest frame using the available radial velocities provided by \citet{laf20}, which were derived from the CARMENES survey spectra. The quality of the resulting template spectra depends on the brightness, number of observations per target and the quality of the telluric correction. The typical values of their S/N are in the 100--2\,000 range \citep[see][]{mar21}. These velocity-corrected template spectra cover 5\,200--17\,100\,{\AA}, corresponding to the visual (VIS) and near-infrared (NIR) channels of the CARMENES instrument. The VIS channel covers 5\,200--9\,600~\AA{} and  has a resolving power of $R$~$=$~$94\,600$, while the NIR channel covers 9\,600--17\,100\,\AA{} with $R$~$=$~$80\,400$.\\

In addition to the spectra and the stellar parameters of the investigated stars, we compiled their kinematical membership to different galactic populations from {\color{blue} Cort{\'e}s-Contreras et al. (in prep)}. This membership was computed from the galactocentric space velocities $U$, $V$, and $W$ as in \citet{mon01} with the methodology described by \citet{uvw87}. We employed the radial velocities provided by \citet{laf20} in combination with the stellar coordinates, parallaxes, and proper motions provided by the {\it Gaia} DR3 \citep{EDR3}. According to the criteria given by \citet{mon18}, they were classified into   different galactic kinematic populations:  the young disc (YD), thin (D), thin-to-thick transition (TD-D), and thick discs (TD). There are no halo stars in our sample. Finally, the stellar parameters, the spectral types (SpTs),  S/N ratios, and  membership to different galactic populations of the stars in our sample are displayed in Table~\ref{tab:sample}.\\ 

\subsection{Abundance determination}
\label{sec:analysis}

We used the spectral synthesis method to compute the abundances of Mg and Si of our target K and M stars. Our analysis employed six~\ion{Mg}{i} lines and six~\ion{Si}{i} lines for which we list the relevant atomic parameters in Table~\ref{tab:MgSilines}. The  spectral synthesis was done using the radiative transfer code {\tt Turbospectrum\footnote{\url{https://github.com/bertrandplez/Turbospectrum2019/}}} \citep{ple12}. We employed the BT-Settl model atmospheres \citep{all12} covering the range from 2\,600 to 4\,500~K in $T_{\rm eff}$ with a step of 100~K, 4.0 to 6.0~dex in $\log{g}$ with a step of 0.5~dex, whereas we took the following metallicity ([Fe/H]) values: $-$1.0, $-$0.5, $+$0.0, $+$0.3, and $+$0.5~dex. We assumed the Solar abundances provided by \citet{asp09}, which are fully-consistent with the abundances used to compute the atmospheric models in the BT-Settl grid. In short, the grid choice covered the metallicities of the studied stars \citep[see e.g.][]{pas18,sch19,mar21}.\\

\begin{table}
\small
\centering
\caption{Atomic parameters for the \ion{Mg}{i} and \ion{Si}{i} lines used in the abundance analysis$^a$.}  
\label{tab:MgSilines}
\begin{tabular}{cccc}
\hline\hline
\noalign{\smallskip}
$\lambda$ & Species & $\chi_l$  & $\log{gf}$\\ 
{[\AA]} &         &   [eV]    &  [dex]   \\
\noalign{\smallskip}
\hline  %\hline
\noalign{\smallskip}
 5711.088  &  \ion{Mg}{i}  &  4.35  & --1.724 \\
 7387.685  &  \ion{Mg}{i}  &  5.61  & --1.000 \\
 7657.603  &  \ion{Mg}{i}  &  5.11  & --1.268 \\
 7659.152  &  \ion{Mg}{i}  &  5.11  & --1.489 \\
 8806.757  &  \ion{Mg}{i}  &  4.35  & --0.134 \\
11828.171  &  \ion{Mg}{i}  &  4.35  & --0.333 \\
\noalign{\smallskip}
10603.425  &  \ion{Si}{i}  &  4.93  & --0.305 \\
10749.378  &  \ion{Si}{i}  &  4.93  & --0.205 \\
10786.849  &  \ion{Si}{i}  &  4.93  & --0.303 \\
10827.088  &  \ion{Si}{i}  &  4.95  & +0.302 \\
10869.536  &  \ion{Si}{i}  &  5.08  & +0.371 \\
12270.692  &  \ion{Si}{i}  &  4.95  & --0.396 \\
\noalign{\smallskip}
\hline
\end{tabular}
\tablefoot{$^a$ $\lambda$: air wavelength, $\chi_l$: excitation potential, $\log{gf}$: oscillator strength.}
\end{table}

In addition to the model atmospheres, we gathered the atomic and molecular data for the spectral synthesis of the \ion{Mg}{i} and \ion{Si}{i} lines from the Vienna Atomic Line Database\footnote{\url{http://vald.astro.uu.see}} \citep[VALD3,][]{rad15}, which we complemented with literature data of the following molecular species: H$_2$O \citep{Bar06}, FeH \citep{Dul03}, MgH \citep{Kur14}, CO \citep{Goo94}, SiH \citep{Kur14}, OH \citep{Kur14}, VO \citep{kem16}, CaH, ZrO, and TiO \citep[B.~Plez priv. comm.; see also][]{Hei21}. This choice of atomic and molecular data is fully consistent with the assumed stellar  parameters \citep{mar21}. We used a $\chi^2$ fitting procedure to reproduce the twelve \ion{Mg}{i} and \ion{Si}{i} lines listed in Table~\ref{tab:MgSilines}. This procedure requires a good prior knowledge of the stellar atmospheric parameters (namely $T_{\rm eff}$, $\log{g}$, [Fe/H], and $\varv \sin{i}$) and leaves the abundance of the line under analysis as a free parameter. These stellar atmospheric parameters were assumed to be the ones computed by \citet[][see also Table~\ref{tab:sample}]{mar21} using the {\sc SteParSyn} code\footnote{\url{https://github.com/hmtabernero/SteParSyn/}} \citep{tab22} while the $\varv \sin{i}$ were computed with the approach of \citet{rei18}. These $\varv \sin{i}$ values were also used by \citet{mar21} to compute the stellar parameters of the target stars adopted in this work.\\ 

To perform the spectral synthesis, we interpolated an stellar atmospheric model using the BT-Settl grid with the parameters of the stars under analysis (i.e. $T_{\rm eff}$, $\log{g}$, and [Fe/H]). The interpolation was carried out in the three-dimensional parameter space covered by the BT-Settl atmospheric models with the routines available in the {\tt SciPy} Python library \citep{scipy}. This interpolated atmospheric model was used with {\tt Turbospectrum} to produce a synthetic spectrum that we later convolved with a rotation kernel to account for the $\varv \sin{i}$ of the star, followed by a Voigt kernel corresponding to the CARMENES line spread function \citep{nag23}. The resulting synthetic spectrum was normalised following the procedure described by \citet{tab22} and  compared to the observed spectrum. We repeated this process until we found an abundance that fitted the observed atomic line under analysis. This procedure was performed on a line-by-line basis for each target star.\\

\begin{table}
\small
\centering
\caption{Abundance sensitivities to the uncertainties in the stellar parameters for two representative stars in our sample. }  
\label{tab:MgSierrors}
\begin{tabular}{lcccc}
\hline\hline
\noalign{\smallskip}
Stellar parameters$^a$ & \multicolumn{2}{c}{J00183+440 (M1.0\,V)}  & \multicolumn{2}{c}{J07274+052 (M3.5\,V)} \\
                      &   $\Delta$[Mg/H] & $\Delta$[Si/H]  & $\Delta$[Mg/H] & $\Delta$[Si/H] \\
                      & [dex] & [dex] & [dex] & [dex] \\
\noalign{\smallskip}
\hline  
\noalign{\smallskip}
 $T_{\rm eff}$~$\pm$~72~K  &  $\mp$~0.06 & $\mp$~0.14  &  $\mp$~0.07  & $\mp$~0.13 \\
 $\log{g}$~$\pm$~0.09~dex &  $\pm$~0.01 &  $\pm$~0.05  &  $\pm$~0.01  &  $\pm$~0.03 \\
 $[$Fe/H$]$~$\pm$~0.10~dex &  $\mp$~0.02 &  $\mp$~0.02  &  $\pm$~0.01  &  $\pm$~0.04 \\
 $\varv \sin{i}$~$\pm$~1.0~km~s$^{-1}$ & $\pm$~0.02 &  $\pm$~0.02 &  $\pm$0.03 &  $\pm$~0.02  \\
 \noalign{\smallskip}
 \hline
 \noalign{\smallskip}
 total  &  0.07 & 0.15 &  0.08 & 0.14  \\
\hline
\end{tabular}
\tablefoot{$^a$The stellar parameter were taken from \citet{mar21}. The two stars are GX~And (J00183+440) and Lutyen's Star (J07274+052).}
\end{table}

Finally, the abundances of magnesium, [Mg/H], and silicon, [Si/H], were obtained by computing the median of the abundance values measured for the available lines of either Mg or Si. Their uncertainties were computed as the median absolute deviation (MAD) multiplied by a factor of 1.4826 to compute the line-to-line scatter \citep{madfactor}. We provide these abundances and their corresponding line-to-line scatter in Table~\ref{tab:sample}. In addition, errors in either [Mg/H] and [Si/H] arising from uncertainties in the stellar atmospheric parameters were calculated for two representative stars. To derive these errors, we varied the $T_{\rm eff}$, $\log{g}$, and [Fe/H], on a one-by-one basis, within their formally derived systematic uncertainties as given by \citet{mar21}: 72~K, 0.09~dex, and 0.10~dex, respectively. In addition, we assumed a formal error bar on $\varv \sin{i}$ of 1.0~km~s$^{-1}$. Then, the total error bars were computed by summing each source of error in quadrature. The representative total uncertainties in our abundance measurements are in the range 0.07--0.15~dex, and we list them in Table~\ref{tab:MgSierrors}.\\

\section{Abundance results and validation}
\label{sec:abundances}

\begin{figure}[]                                                 
   \centering                       
    \includegraphics[width=0.55\textwidth]{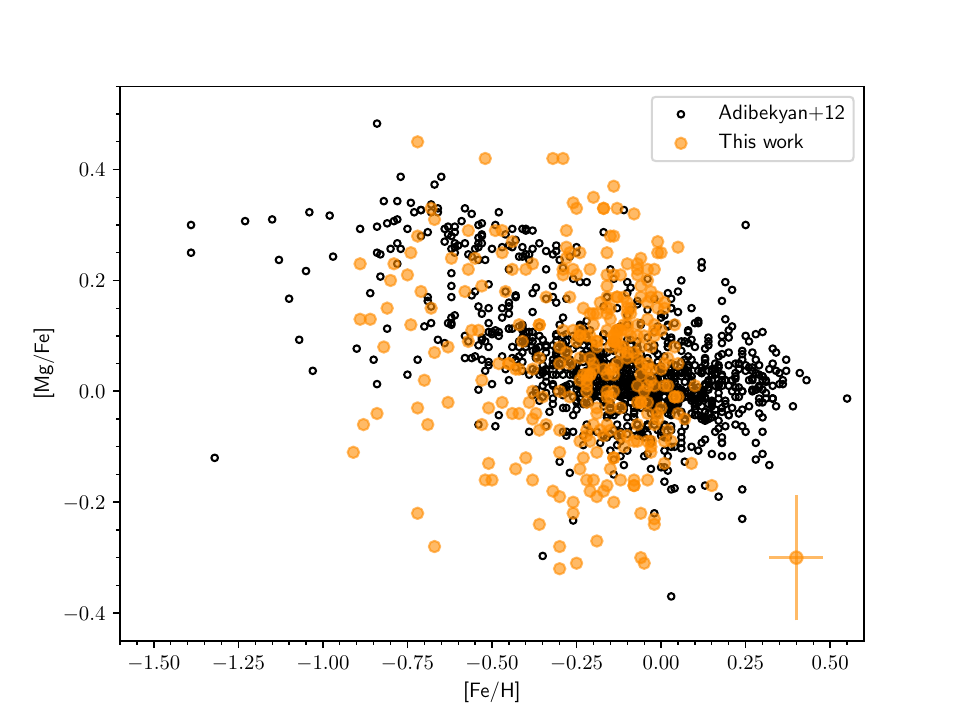}
    \includegraphics[width=0.55\textwidth]{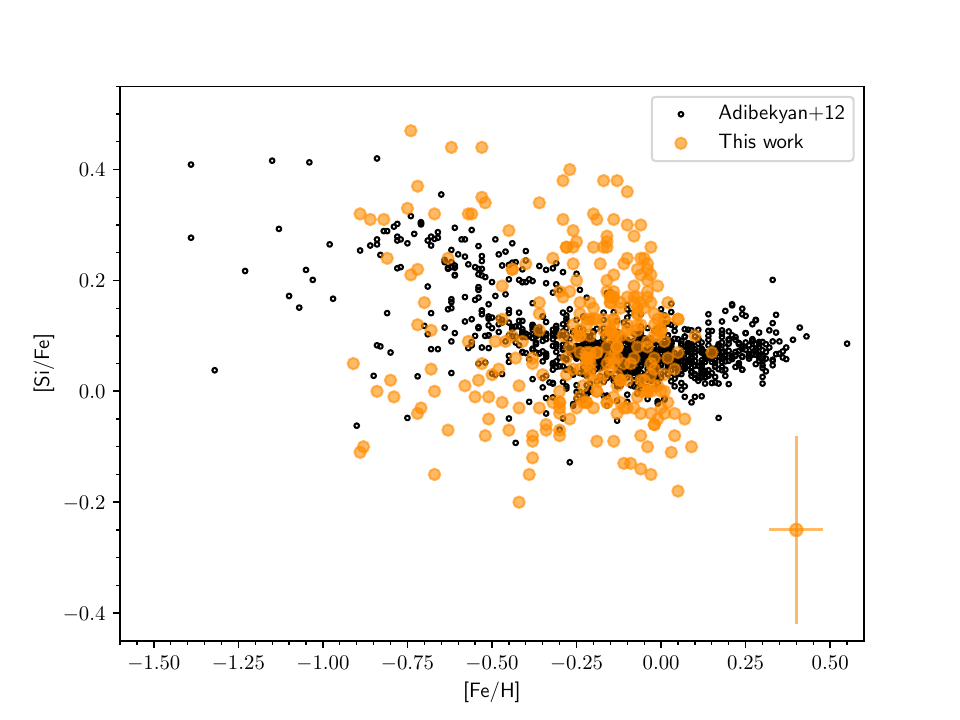}
    \caption{[X/Fe] vs. [Fe/H] for both Mg (top panel) and Si (bottom panel). Our target stars are represented by orange circles while the black open circles correspond to the abundances derived by for 1\,111 FGK stars calculated by \citet{adi12}.  The typical error bar for each abundance is displayed in the bottom right of each panel.}
    \label{fig:abundances}
\end{figure}

  The M dwarf stars are formed from the same material as the FGK-type population of our Galaxy \citep[see e.g.][]{bud21} and, therefore, they should display similar galactic trends. Thus, using the abundances computed in Sect.~\ref{sec:analysis} alongside the iron content [Fe/H] of the stars in our sample, we computed the [Mg/Fe] and [Si/Fe] ratios\footnote{[X/Fe]~$=$~[X/H]~$-$~[Fe/H]}. In parallel, we compiled the [Fe/H], [Mg/Fe], and [Si/Fe] values for 1\,111 FGK stars computed by \citet{adi12} and plotted them together with our data in Fig.~\ref{fig:abundances}. Overall, we found good agreement across the $T_{\rm eff}$ domain covered by our sample for both Mg and Si. For Mg, the scatter is somewhat larger than that of the FGK stars (Fig.~\ref{fig:abundances}, top panel), while the Si abundances show a larger scatter (Fig.~\ref{fig:abundances}, bottom panel).  A compatible behaviour with the FGK stars has also been found for other chemical elements in M dwarfs observed with CARMENES. In fact, the chemical abundances of the heavy elements Rb, Sr, and Zr were computed by \citet{abi20} and found to be in agreement with the abundances of stars of higher temperatures in our galaxy.  V was also studied by \citet{sha21} using stars in the analysed data set. Moreover, the sequences formed by the different elements should respond to the chemical composition (and its evolution) of the stars in our Galaxy. This agreement between the trends formed by Mg and Si for M dwarfs and their FGK counterparts was also studied by \citet{ish20,ish22}. \\

The \ion{Mg}{i} and \ion{Si}{i} lines in the spectra of M dwarfs later than M5.5V  could not be fitted by our method, yielding extreme nonphysical values that are due to the contamination of our lines by neighbouring features (due to e.g. TiO, VO, ZrO, ...). In addition, the methodology used here is not suited to model the spectra of late M dwarfs, due to the formation of dust in their atmospheres \citep{tsu96,all01} which was not considered in the version of the BT-Settl models employed in this work. Moreover, a similar SpT limit was already reported by \citet{raj18} and tied to dust condensation processes. Finally, the onset of dust condensation in late M dwarfs responds to  various physical and chemical processes that complicate our current understanding of their spectra, thus making determination of Mg and Si abundances at SpTs M6.0~V or later a challenging task that is beyond the scope of the methodology employed in this work. 

\subsection{Multiple systems}
\label{sec:binary}

Since multiple systems are assumed to be born from the same parental cloud and are expected to be coeval, and under that assumption, the chemical composition of each star in the system should be the same \citep[see e.g.][]{brew16}. The sample studied in this work contains stars that belong to wide multiple stellar systems. Among our sample, seven stars studied here are companions to an FGK star and they were analysed by \citet{mon18}, while a dozen M companions to yet another M star are also analysed in this work using only CARMENES spectra.  As the abundances of the former stars were computed independently, they can provide valuable information on the accuracy of the abundance analysis of the present work. We provide relevant information regarding these multiple systems in Table~\ref{tab:binary}.\\

\begin{figure}[]                                                 
   \centering                       
    \includegraphics[width=0.55\textwidth]{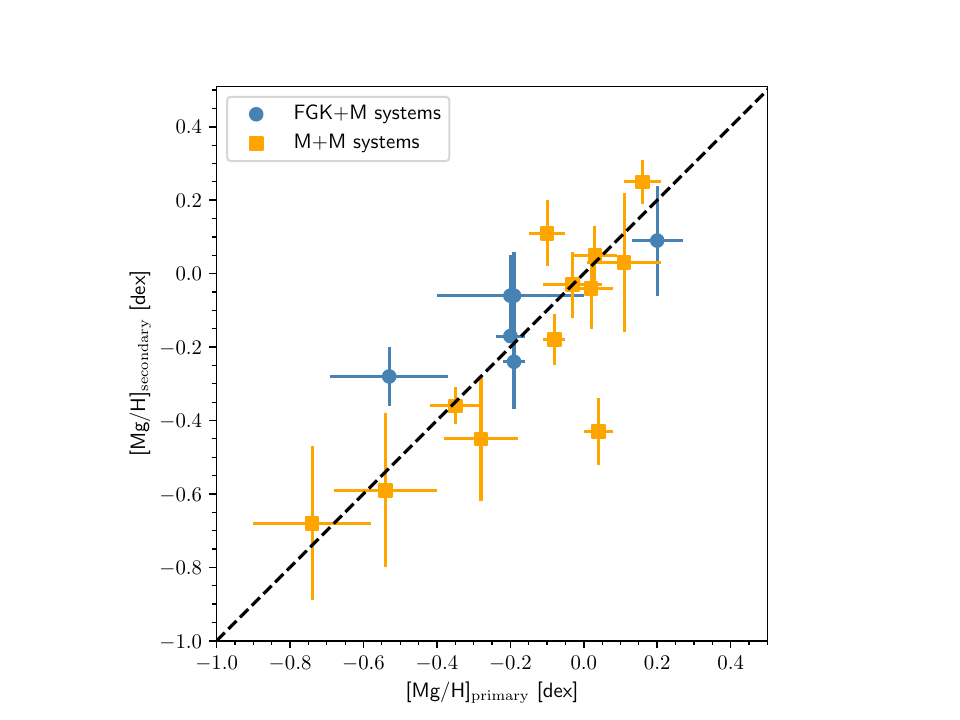}
    \includegraphics[width=0.55\textwidth]{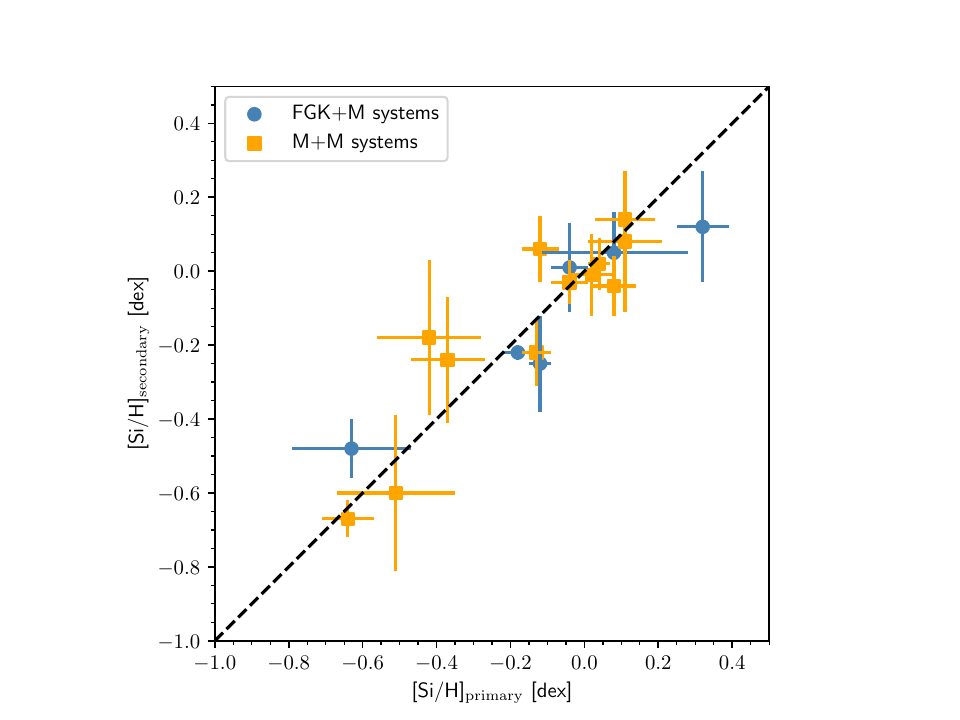}
    \caption{Comparison in both [Mg/H] (top panel) and [Si/H] (bottom panel) between the components in FGK$+$M (blue) and M$+$M (orange) systems.}

\label{fig:binary}
\end{figure}

A comparison between the derived abundances for each component is displayed in Fig.~\ref{fig:binary}. Our method provides abundances that are similar for both components with a dispersion of 0.10~dex (for Mg) and 0.12~dex (for Si) for the M dwarfs that are companions to an FGK primary, while the systems with only two M stars show a dispersion of 0.11~dex (Mg) and 0.16~dex (Si). If we combine both samples the total scatter is 0.11~dex for Mg and 0.16~dex for Si. The dispersion is smaller for systems with an FGK primary indicating that the intrinsic abundance scatter of the abundances of FGK stars is smaller than for the M stars (see Fig.~\ref{fig:abundances}). The results displayed in Fig.~\ref{fig:binary} show good agreement between the abundances of both primary and secondary stars. This estimation of the error bars is on or below the uncertainty level for previous studies computing both Mg and Si for M dwarfs. For comparison, \citet{ish20} estimated that the [Mg/H] error bar for an M dwarf is in the range of 0.2--0.3~dex. Later on, \citet{ish22} estimated an error bar for M dwarfs at the level of 0.3~dex for Mg and Si. \citet{mal20} provided an error bar at the level of 0.15~dex. However, their Mg abundance pattern is about $-$0.2~dex offset from that of FGK stars, which might suggest a significant systematic error. Finally, the scatter for our targets in multiple systems is at the 0.1--0.2~dex level, which is equal to or better than previous studies of the abundances of Mg and Si for M dwarfs. In turn, these uncertainties are also at the level of the error bars inferred in Sect.~\ref{sec:analysis}. The dispersion found for these multiple systems can be due diffusion processes that occur in solar-like stars \citep[see e.g.][]{tal05,kor07} or even planet engulfment \citep[e.g. ][]{spi21}. These processes can result in different photospheric abundances for stars belonging to the same multiple system. According to the literature this difference can be up to $\approx$1.0~dex for a 1.4$M_\odot$ star, and up to $\approx$~0.1~dex for a 1~M$_\odot$ star \citep[see ][]{dea18}. However,  the earliest star in the multiple systems studied here is $\rho$~Cnc~A (G8.0\,V) and a $T_{\rm eff}$ of $\approx$~5299~K that according to the models of \citet{dea18} can result in difference of $\approx$~0.1 dex, which is compatible with the scatter found for FGK$+$M systems. Finally, in Fig.~\ref{fig:binary} there are two outliers with respect the 1:1 trend. The first of them is the system V388~Cas (M1.5\,V) and BD$+$61 195 (M5.0\,V)  and it deviates only for Mg while the second system is $\rho$~Cnc~A (G8.0\,V) and $\rho$~Cnc~B (M4.5\,V). Interestingly, we also note that BD$+$61~195 has a tenative earth-like exoplanet \citep{per19} and the star $\rho$~Cnc~A has five known exoplanets \citep{bou18}. Thus, this systems should be explored for more exoplanets in order to constrain planet formation theories. \\ 

\begin{table*}
\centering
\caption{Stellar parameters ($T_{\rm eff}$, $\log{g}$, [Fe/H]) and chemical abundances ([Mg/H], [Si/H]) for the resolved multiple systems with M stars.}

\label{tab:binary}
\begin{tabular}{lllccccc}
\hline\hline
\noalign{\smallskip}
Name  & Karmn & SpT &   $T_{\rm eff}$ & $\log{g}$ & [Fe/H]  & [Mg/H] & [Si/H]\\
      &       &     &   [K]           & [dex]     & [dex]   &  [dex] & [dex]\\
      \noalign{\smallskip}
\hline
\noalign{\smallskip}
V538~Aur         &  \ldots        & K1\,V   & $5292\pm32$ & $4.38\pm0.09$ &  $0.04\pm0.02$  & $-0.20\pm0.20$	& $0.08\pm0.05$\\
HD~233153        & J05415$+$534   & M1.0\,V & $3825\pm14$ & $4.94\pm0.10$ & $-0.02\pm0.06$  & $-0.06\pm0.11$    & $0.05\pm0.02$\\
\noalign{\smallskip}
V869~Mon         & \ldots         & K0/2\,V & $4918\pm63$ & $4.36\pm0.06$ & $-0.11\pm0.03$  & $-0.19\pm0.05$ &	$-0.04\pm0.05$   \\
GJ~282~C         & J07361$-$031   & M1.0\,V & $3825\pm12$ & $5.06\pm0.08$ & $-0.04\pm0.06$  & $-0.06\pm0.12$ & $0.01\pm0.02$ \\
\noalign{\smallskip}
HD~154363~A	     & \ldots         & K5\,V   & $4710\pm103$& $4.40\pm0.38$ & $-0.62\pm0.05$  & $-0.53\pm0.16$	& $-0.63\pm0.13$  \\
HD~154363~B      & J17052$-$050   & M1.5\,V & $3587\pm14$ & $4.89\pm0.10$ & $-0.39\pm0.07$  & $-0.28\pm0.08$ & $-0.48\pm0.02$\\
\noalign{\smallskip}
HD~16160         & \ldots         & K3\,V   & $4831\pm59$ &	$4.24\pm0.19$ & $-0.20\pm0.02$  & $-0.19\pm0.03$	& $-0.12\pm0.06$ \\
BX~Cet           & J02362$+$068   & M4.0\,V & $3335\pm45$ & $4.91\pm0.10$ & $-0.24\pm0.12$  & $-0.24\pm0.13$ &  $-0.25\pm0.22$\\
\noalign{\smallskip}
$o^{02}$~Eri~A   & \ldots         & K0.5\,V & $5128\pm31$ & $4.37\pm0.08$ & $-0.37\pm0.02$  & $-0.20\pm0.04$ & $-0.18\pm0.04$\\
$o^{02}$~Eri~C   & J04153$-$076   & M4.5\,V & $3179\pm61$ & $5.00\pm0.18$ & $-0.30\pm0.17$  & $-0.17\pm0.02$ & $-0.22\pm0.26$\\
\noalign{\smallskip}
$\rho$~Cnc~A     & \ldots         & G8.0\,V   & $5299\pm58$ &	$4.35\pm0.13$ & $0.29\pm0.04$  & $0.20~\pm~0.07$	& $0.32\pm0.05$ \\
$\rho$~Cnc~B     & J08526$+$283   & M4.5\,V & $3321\pm37$ & $4.87\pm0.08$ & $-0.10\pm0.11$ & $0.09\pm0.15$ & $0.12\pm0.23$\\
\noalign{\smallskip}
GX~And           & J00183$+$440   & M1.0\,V & $3603\pm24$ & $4.99\pm0.14$ & $-0.52\pm0.11$ & $-0.54\pm0.14$ & $-0.42\pm0.02$ \\
GQ~And           & J00184$+$440   & M3.5\,V & $3318\pm53$ & $5.20\pm0.11$ & $-0.36\pm0.17$ & $-0.59\pm0.21$ & $-0.18\pm0.09$ \\
\noalign{\smallskip}
BD+61~195 & J01026$+$623          & M1.5\,V & $3791\pm19$ & $4.76\pm0.11$ &  $0.05\pm0.06$ & $0.04\pm0.04$ & $-0.13\pm0.02$\\
V388~Cas  & J01033$+$623          & M5.0\,V & $3057\pm49$ & $5.12\pm0.18$ & $-0.24\pm0.24$ & $-0.43\pm0.09$ & $-0.22\pm0.04$ \\
\noalign{\smallskip}
PM~J02489$-$1432W  & J02489$-$145W   & M2.0\,V & $3655\pm25$ & $4.98\pm0.10$ &  $0.04\pm0.05$ & $0.16\pm0.05$ & $-0.04\pm0.07$ \\
PM~J02489$-$1432E  & J02489$-$145E   & M2.5\,V & $3572\pm27$ & $4.94\pm0.12$ &  $0.00\pm0.04$ & $0.25\pm0.06$ & $-0.03\pm0.03$ \\
\noalign{\smallskip}
V2689~Ori          & J05365$+$113   & M0.0\,V & $4067\pm14$ & $5.04\pm0.07$ &  $0.01\pm0.03$ & $-0.08\pm0.03$ & $0.04\pm0.02$ \\
PM~J05366$+$1117   & J05366$+$112   & M4.0\,V & $3355\pm23$ & $5.17\pm0.16$ & $-0.20\pm0.10$ & $-0.18\pm0.07$ & $0.02\pm0.16$ \\
\noalign{\smallskip}
HD~79210           & J09143$+$526   & M0.0\,V & $4015\pm16$ & $4.91\pm0.08$ & $-0.12\pm0.05$ & $-0.10\pm0.05$ & $-0.12\pm0.02$\\
HD~79211           & J09144$+$526   & M0.0\,V & $3983\pm12$ & $5.17\pm0.07$ & $-0.03\pm0.04$ & $0.11\pm0.09$ & $0.06\pm0.02$ \\
\noalign{\smallskip}
GJ~360             & J09425$+$700   & M2.0\,V & $3547\pm23$ & $5.02\pm0.12$ &  $0.00\pm0.03$ & $0.11\pm0.10$ & $0.11\pm0.05$ \\
GJ~362             & J09428$+$700   & M3.0\,V & $3504\pm30$ & $5.06\pm0.12$ &  $0.02\pm0.05$ & $0.03\pm0.19$ & $0.08\pm0.12$ \\
\noalign{\smallskip}
BD+44~2051~A       & J11054$+$435   & M1.0\,V & $3628\pm19$ & $5.01\pm0.13$ & $-0.56\pm0.09$ & $-0.74\pm0.16$ & $-0.51\pm0.02$ \\
WX~UMa             & J11055$+$435   & M5.5\,V & $3278\pm86$ & $5.25\pm0.20$ & $-0.36\pm0.25$ & $-0.68\pm0.21$ & $-0.60\pm0.23$ \\
\noalign{\smallskip}
HD~97101~A         & J11110$+$304E  & K7.0\,V & $4211\pm13$ & $4.98\pm0.07$ &  $0.04\pm0.03$ & $0.03\pm0.06$ & $0.08\pm0.02$ \\
HD~97101~B         & J11110$+$304W  & M2.0\,V & $3730\pm20$ & $4.78\pm0.13$ & $-0.05\pm0.07$ & $0.05\pm0.08$ & $-0.04\pm0.02$ \\
\noalign{\smallskip}
BD+24~2733~A       & J14257$+$236W  & M0.0\,V & $3985\pm13$ & $4.89\pm0.08$ &  $0.07\pm0.04$ & $0.02\pm0.06$ & $0.02\pm0.02$\\
BD+24~2733~B       & J14257$+$236E  & M0.5\,V & $3933\pm12$ & $4.71\pm0.11$ &  $0.09\pm0.04$ & $-0.04\pm0.11$ & $-0.01\pm0.04$ \\
\noalign{\smallskip}
HD~147379          & J16167$+$672S  &  M0.0\,V &  $4034\pm17$ & $4.78\pm0.09$ &  $0.00\pm0.04$ & $-0.03\pm0.08$ & $0.11\pm0.04$ \\
EW~Dra             & J16167$+$672N  &  M3.0\,V &  $3569\pm32$ & $4.97\pm0.11$ & $-0.02\pm0.06$ & $-0.03\pm0.09$ & $0.14\pm0.06$ \\
\noalign{\smallskip}
HD~173739          & J18427$+$596N  & M3.0\,V &  $3473\pm34$ & $4.90\pm0.11$ & $-0.31\pm0.12$ & $-0.28\pm0.10$ & $-0.37\pm0.06$ \\
HD~173740          & J18427$+$596S  & M3.5\,V &  $3493\pm48$ & $4.98\pm0.12$ & $-0.38\pm0.18$ & $-0.45\pm0.17$ & $-0.24\pm0.06$ \\
\noalign{\smallskip}
Ross~730           & J19070$+$208   &  M2.0\,V &  $3543\pm21$ & $5.03\pm0.12$ & $-0.46\pm0.07$ & $-0.35\pm0.05$ & $-0.64\pm0.07$\\
HD~349726          & J19072$+$208   &  M2.0\,V & $3558\pm19$ & $5.06\pm0.10$ & $-0.46\pm0.06$  & $-0.36\pm0.07$ & $-0.67\pm0.07$ \\
\noalign{\smallskip}
\hline
\noalign{\smallskip}
\end{tabular}
\end{table*}

\subsection{Trends with galactic kinematics}
\label{sec:kinematics}
\begin{figure}[ht]                                                 
   \centering       
   \includegraphics[width=0.5\textwidth]{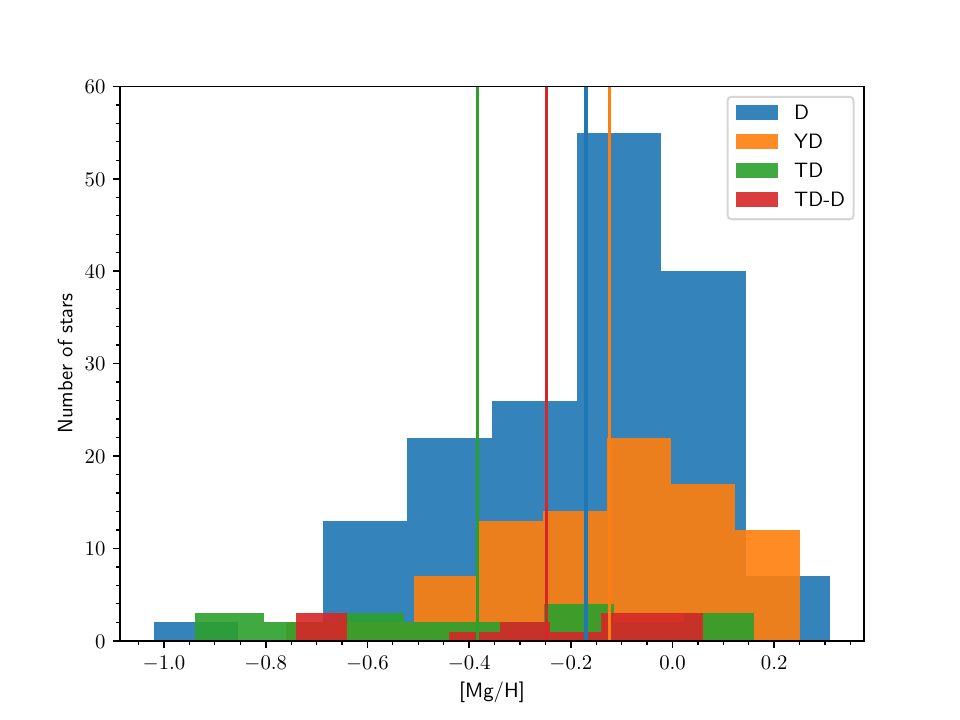}
    \includegraphics[width=0.5\textwidth]{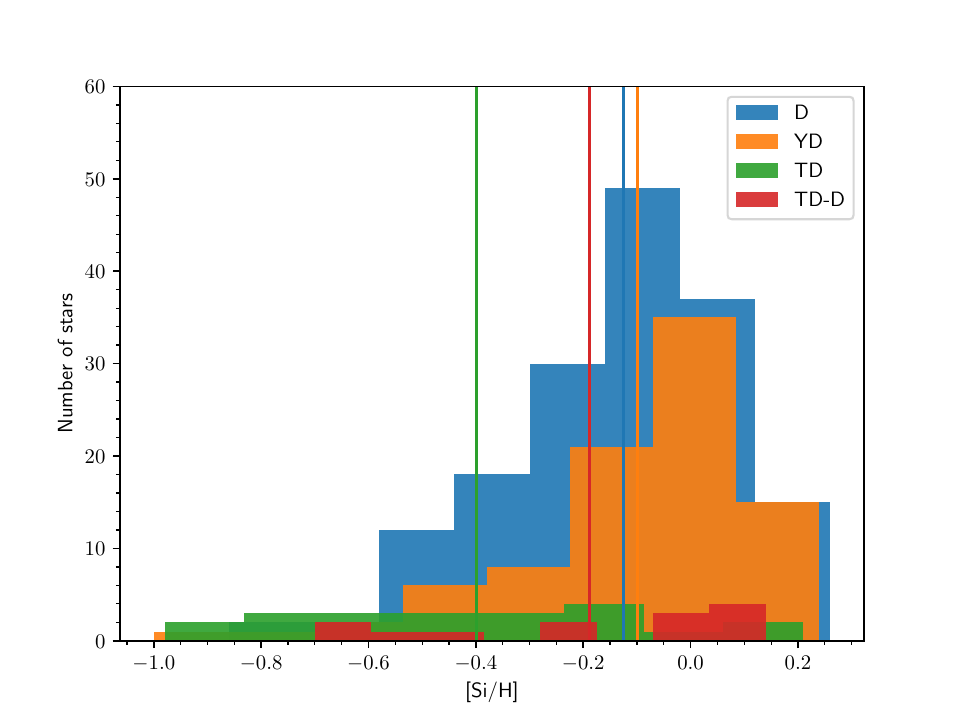}
    \caption{M dwarf abundance distributions for Mg and Si separated according to the kinematic
membership of stars in the thick disc (TD), thick disc-thin disc
transition (TD-D), thin disc (D), and young disc (YD), following the criteria of \citet{mon18} (see text for details). Mean values for each population are indicated by vertical lines.} 
\label{fig:pops}
\end{figure}

The stars analysed in this work can be classified into four different kinematic categories described in Sect.~\ref{sec:dataproc} (YD, D, TD-D, and TD). Using the kinematic membership of the analysed CARMENES stars \citep[see e.g. ][]{sha21, mar21}, we checked if the populations show a different chemical abundances for either Si or Mg. We display this kinematic classification in Fig.~\ref{fig:pops} and Table~\ref{tab:sample}. A similar analysis was done by \citet{sha21} for the V abundances and by \citet{mar21} for the Fe abundances of the M dwarfs. \citet{sha21} found no meaningful trend with [V/Fe] for the stars belonging to the thick or thin disc. However, \citet{mar21} found that the mean metallicity of M dwarfs belonging to the thin disc is lower than that of those belonging to the thick disc, in agreement with previous works \citep{ben03,ben05}.\\ 

We display the mean values of our [Mg/H] and [Si/H] values in Fig.~\ref{fig:pops} and Table~\ref{tab:pops}, which reveals decreasing abundances towards older kinematic components of the Galaxy. That is, stars belonging to the thick and transition discs (TD, TD-D) tend to have lower Mg and Si abundances than those belonging to the thin disc (YD, D). This is in line with the iron abundances presented by \citet{mar21} and in agreement with the expected trend. Furthermore, studies based on FGK stars show that [Mg/Fe] and [Si/Fe] are distinct between thin and thick-disc populations, with the stars of the thick disc having on average higher [Mg/Fe] and [Si/Fe] values due to $\alpha$-enrichment \citep{adi12,san15}. Consequently, the type of planets that can be around stars belonging to the different galactic populations could be systematically different. Therefore, kinematical information could be key to predicting and interpreting these planets \citep[see e.g.][]{bit20}.

\begin{table}
\small
\centering
\caption{Mean abundance values for each kinematic population (see Sect.~\ref{sec:kinematics} for details).}  
\label{tab:pops}
\begin{tabular}{lcc}
\hline\hline
\noalign{\smallskip}
Population & [Mg/H]  & [Si/H] \\
\noalign{\smallskip}
\hline  
\noalign{\smallskip}
YD   & $-$0.12~$\pm~$0.22 & $-$0.09~$\pm$~0.23 \\
D    & $-$0.18~$\pm$~0.24 & $-$0.13~$\pm$~0.23 \\
TD-D & $-$0.24~$\pm$~0.26 & $-$0.19~$\pm$~0.32 \\
TD   & $-$0.38~$\pm$~0.35 & $-$0.40~$\pm$~0.31\\
 \noalign{\smallskip}
 \hline
\end{tabular}
\end{table}
\section{Implication for planets}

\label{sec:imp}

\subsection{Planetary building blocks and constraints on planetary structure}
\label{sec:ratios}

Mineral compounds form the crust, mantle, and core of rocky planets. The abundances of Mg, Si, and Fe provide valuable information on these compounds and can be used to model the internal structure of rocky planets that can be formed in a given planetary system \citep[see e.g.][and references therein]{adi21}.  Using the approach described by \citet{san15,san17}, we considered the mass fraction of the mineral compounds predicted by equilibrium condensation models \citep{lod03,sea07}.  Thus, we used the photospheric abundances of the refractory elements Si, Mg, and Fe as input, we computed the relative mass for the following mineral compounds: Fe, MgSiO$_3$, Mg$_2$SiO$_4$, and SiO$_2$ with simple stoichiometry, as discussed by  \citet{san15}. From these relative masses for the mineral compounds, we inferred the iron-to-silicate ratio ($f_{\rm iron}$) for a given set of stellar abundances. We computed $f_{\rm iron}$ for our target stars, using the Mg and Si abundances derived in Sect~\ref{sec:analysis} alongside the Fe abundances derived by \citet{mar21}. In all, $f_{\rm iron}$ was given by \citet{san15,san17} as:\\

\begin{equation}
\label{eq:firondef}
f_{\rm iron} = \frac{m_{\rm Fe}}{m_{\rm Fe} + m_{\rm MgSiO_3} + m_{\rm Mg_2SiO_4} + m_{\rm SiO_2}}
\end{equation}

\noindent where m$_X$~=~N$_X$~$\mu_X$, N$_X$ represents the number of particles for species X, and $\mu_X$ corresponds to the molecular weight of species X.

\begin{figure}                                                 
   \centering    
    \includegraphics[width=0.45\textwidth]{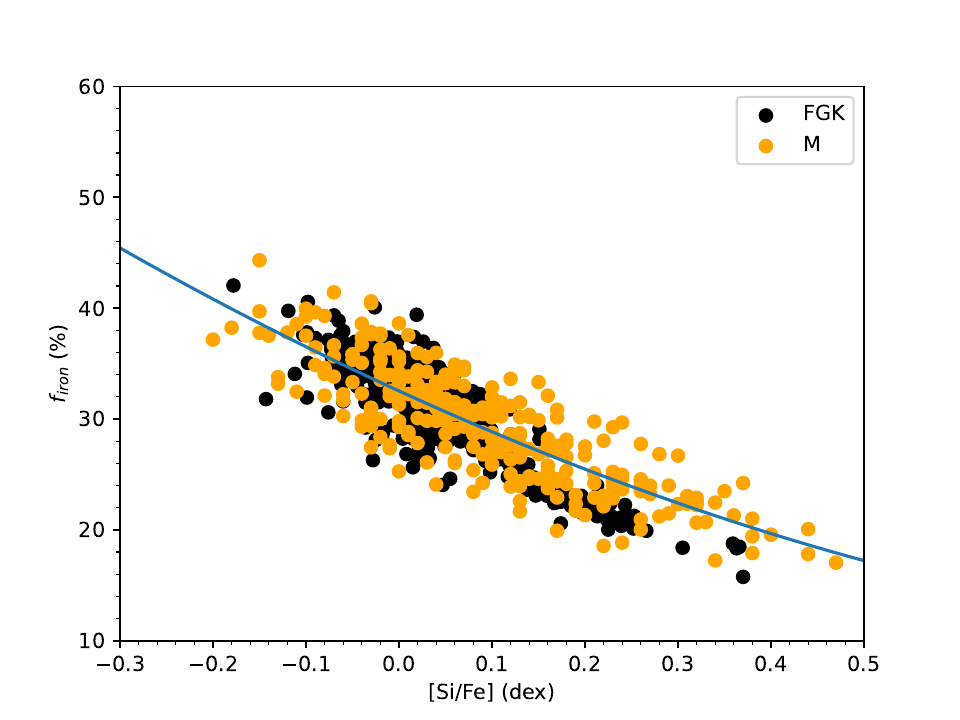}
    \includegraphics[width=0.45\textwidth]{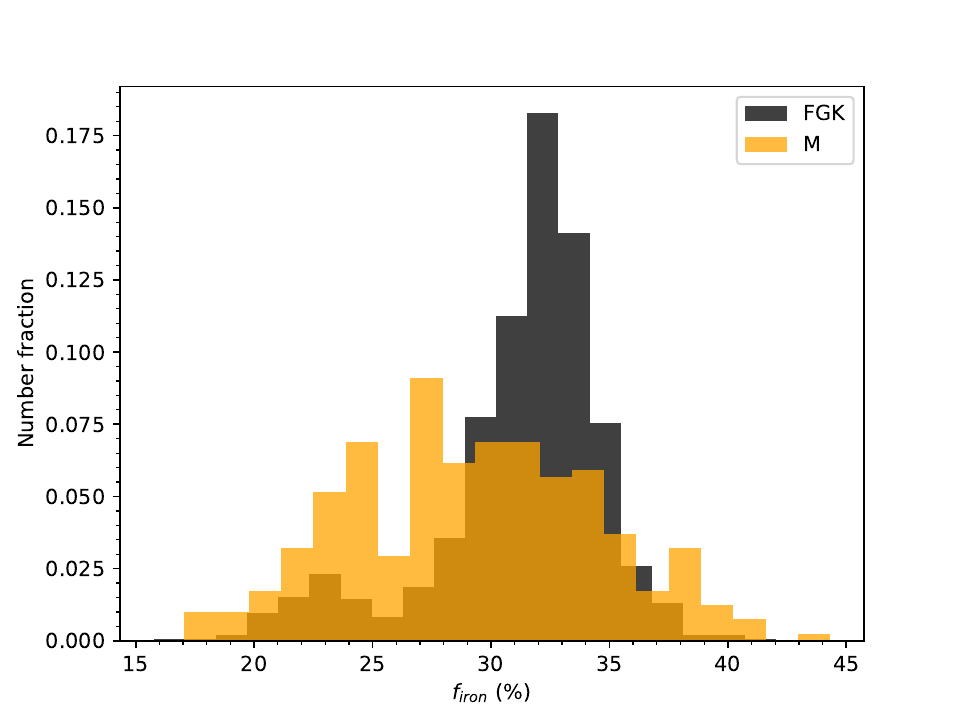}
    \caption{Top panel: [Si/Fe] vs. $f_{\rm iron}$. The target stars are in orange while the black points correspond to the $f_{\rm iron}$ values computed from the \citet{adi12} abundances. The fit through the M dwarf values is denoted with a blue line. Bottom panel: Histogram for the $f_{\rm iron}$ values of our sample.} 
    \label{fig:firon1}
\end{figure}

 We represent our derived $f_{\rm iron}$ values in Fig.~\ref{fig:firon1}. In these two figures, we compare our values to those of the sample of  FGK stars investigated by \citet{adi12} and \citet{san15}. Furthermore, the values of $f_{\rm iron}$ for our targets are spread over in a range similar to that of the FGK stars. In addition, the FGK distribution is bimodal with peaks at $\approx$~23~\% and $\approx$~33~\%, which is narrower compared to the distribution of our analysed M-dwarfs. They show a distribution with a peak at $f_{\rm iron}$ at $\approx$~29\% and there is a small gap around $f_{\rm iron}$~$\approx$~25\%. Similarly, the gap appears to be related to the bimodality of the $f_{\rm iron}$ distribution of the FGK stars. The distribution of the $f_{\rm iron}$ values suggests that the M dwarfs can form rocky planets similar to those of the FGK stars, although our results also suggest that M stars may produce planets with an $f_{\rm iron}$ value that is on average smaller than for the FGK stars.  However, the larger spread seen in the bottom panel Fig.~\ref{fig:firon1} might be tied to the abundance spread being larger for the M dwarfs (see Sect~\ref{sec:binary}). This might translate into smaller planetary cores \citep{san15,san17}.
 
 The sample analysed in this work can be used to predict the $f_{\rm iron}$ values from the silicon abundances.  In Fig.~\ref{fig:firon1}, we show that the values of [Si/Fe] and $f_{\rm iron}$ are spread similarly regardless of the SpT of the star, although the low $f_{\rm iron}$ and high [Si/Fe] portion of the diagram seems more populated for our sample of M dwarfs.  We fitted a polynomial relationship between [Si/Fe] and $f_{\rm iron}$ following \citet{san17}, which yielded:
 \begin{equation}
\label{eq:fironsi}
f_{\rm iron} = a - b~{\rm [Si/Fe]} + c{\rm [Si/Fe]}^2
\end{equation}
\noindent where the coefficients take the following values: $a$~$=$~15.5~$\pm$~4.6, $b$~$=$~(38.3~$\pm$~1.8), and $c$~$=$~32.53~$\pm$~0.21. This second order polynomial can be used to predict $f_{\rm iron}$ solely from [Si/Fe] values determined from spectroscopy.

\subsection{Correlation with planet occurence}
\label{sec:withwopla}

\begin{figure*}[]                                                 
   \centering                       
    \includegraphics[width=\textwidth]{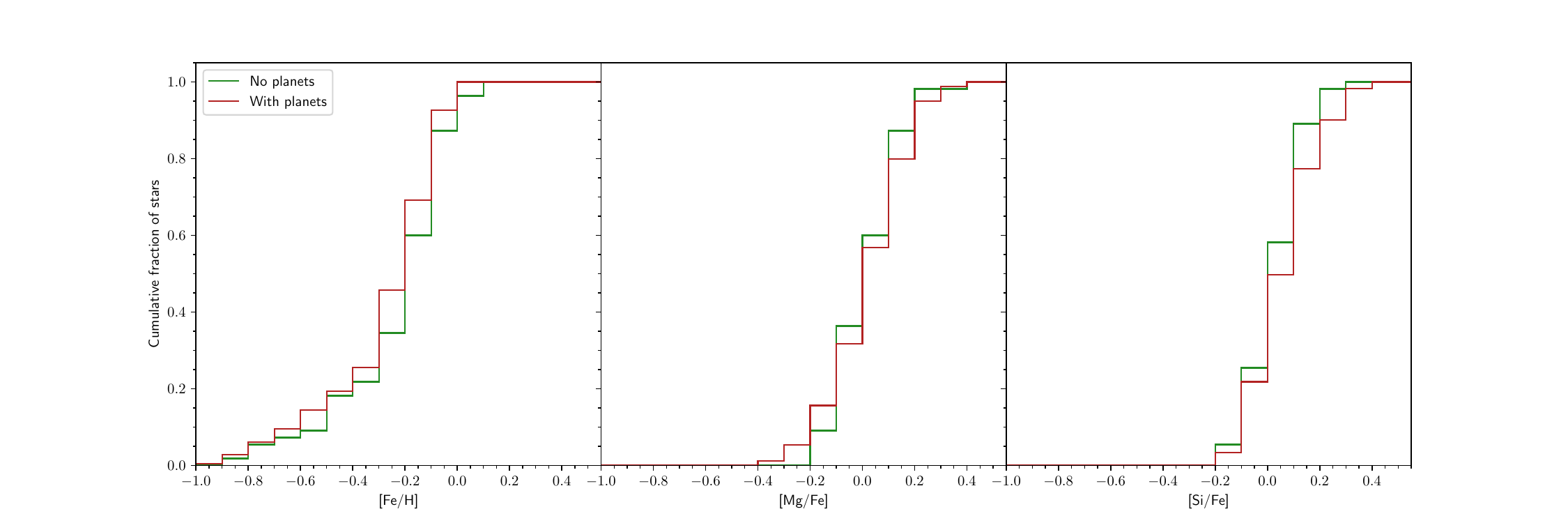}
    \caption{Normalised cumulative histograms for [Fe/H], [Mg/Fe], and [Si/Fe] for the stars analysed in this work. The stars have been separated into two populations: stars with known planets (red) and stars without known planets (green). Our statistical analysis reveals that the distributions corresponding to this two populations are not different according to the Kolmogorov-Smirnoff test (see text for details).}
    \label{fig:wplanets}
\end{figure*}

The probability of hosting a giant exoplanet tends to increase with the metallicity of F-, G-, and K- dwarf stars as shown by many previous works, for example: \citet{gon97}, \citet{san01}, \citet{fisval05}, \citet{gon06}, \citet{gui06}, \citet{ghe10}, \citet{joh10}, \citet{sou11}, \citet{ref15}, and \citet{buc18}. However, this relationship has not been found  for M dwarf stars \citep[e.g.][]{lau04}. In order to investigate the existence of such relation, we searched for confirmed exoplanet discoveries around our targets summarised by \citet{rib23}, and tabulated by the NASA exoplanet archive. These identified planetary systems are listed in Table~\ref{tab:sample}. We then used this information and compared the differences between abundances of stars with and without reported exoplanets.\\

We performed a Kolmogorov-Smirnoff test to assess the difference between the stars with and without planets in our sample. To do this test, we compared the cumulative histograms for [Fe/H], [Mg/Fe], and [Si/Fe] for two different samples (targets with and without planets), we represent these cumulative histograms in Fig.~\ref{fig:wplanets}. Then, we applied a K-S test to these two distributions and found that at the 95\% confidence level Mg, Si, and Fe do not significantly differ for stars with and without planets. In addition, to test the robustness we performed 10\,000 Monte Carlo simulations taking into account the uncertainties of the abundances and applied the K-S test for each element (Mg, Si, and Fe). We found only 1\% of these simulations for Mg show a difference in abundance for stars with and without planets. Si and Fe show a difference only in a 18\% and 13\% of these simulations, respectively. This numbers are indicative of relatively small probability of a differences for stars with and without planets.  This is in contrast with previous results reported for FGK stars. For instance, \citet{mon18} reported a correlation between iron abundance and the presence of planets, while \citet{adi12} only reported a correlation between Mg, Si and planet occurrence only for FGK stars hosting giant planets. However, our findings are in agreement with the results of \citet{mal20} for M dwarf stars, who found no differences in the composition of Mg and Si for M stars with and without planets. The different results found by \citet{mon18} for FGK$+$M systems arise from the fact that only seven of our targets in FGK$+$M systems have been searched for planets by the CARMENES surveys. In fact, according to \citet{rib23}, none of these seven M stars has a known planetary companion.  Since these binary systems are assumed to be coeval, and in consequence are formed from the same parental cloud, they are expected to have the same chemical makeup. Consequently, the presence or absence of planets around either star in a given multiple system could be due to the fact that are still unknown planets  around them. In this regard, stars in these multiple systems should be targeted for additional planets. Exploring the presence of planets around multiple systems is also interesting to study planetary formation around stars with the same metallicity and to constrain planet formation theories since the planets discovered around M dwarfs are mostly rocky \citep[see, e.g.][]{sab21,die23,rib23}. 

\section{Summary and conclusions}
\label{sec:conclusions}

We  computed the magnesium and silicon abundances of 314 K- and M-type dwarfs observed within the framework of the CARMENES survey. Our analysis used the spectral synthesis method to reproduce the data. Specifically, we used the radiative code {\tt Turbospectrum}, a grid of BT-Settl atmosphere models, and a carefully chosen set of atomic and molecular data. Our analysis encompassed of six \ion{Mg}{i} and six \ion{Si}{i} lines at both optical and infrared wavelengths. The method employed here can provide abundances for M stars down to spectral type M5.5\,V. In addition, we compared our derived abundances to the known galactic trends of FGK stars, computed the type of minerals that each analysed M dwarf can form and concluded that both FGK and M dwarf stars have a similar distribution on $f_{\rm iron}$. Finally, the main conclusions of the present work can be summarised as follows:

\begin{enumerate}

\item[--] We compared our results for galactic trends of [Mg/Fe] and [Si/Fe] versus [Fe/H] with the values derived by \citet{adi12}, and found that both trends are compatible within error bars. Therefore, the M dwarfs have the same chemical composition as the solar-like FGK stars.

\item[--] We compared the abundances of Mg and Si derived by \citet{mon18} to the seven M companions to FGK stars in the CARMENES sample. From this comparison, we found that our Mg and Si abundances are consistent at the 0.1--0.2~dex level for the resolved binary systems observed in the framework of the CARMENES consortium.

\item[--] We computed the $f_{\rm iron}$ values as \citet{san17} and found that the M dwarfs should produce (on average) similar rocky planets to that of the FGK stars.

\item[--] We compared the abundances of stars with and without planets and found that no clear statistical connection exists between the abundances an the incidence of planets. 

\item[--] Stars in multiple systems should be targeted in the search additional planets, when posible, in order to constrain planet formation theories and posible connections with the metallicity the presence/absence of planetary companions.

\item[--] The different kinematic populations have different abundances (see Table~\ref{tab:pops}), and they should be explored further by  radial velocity searches of exoplanets around M dwarfs although at present and according to our findings, metallicity does not seem to correlate with the presence of planets, at least for M dwarf stars in the solar neighbourhood. 

\end{enumerate}

\begin{acknowledgements}

 This publication was based on observations collected under the CARMENES Legacy+ project. CARMENES is an instrument at the Centro Astron\'omico Hispano en Andaluc\'ia (CAHA) at Calar Alto (Almer\'{\i}a, Spain), operated jointly by the Junta de Andaluc\'ia and the Instituto de Astrof\'isica de Andaluc\'ia (CSIC).

  CARMENES was funded by the Max-Planck-Gesellschaft (MPG), 
  the Consejo Superior de Investigaciones Cient\'{\i}ficas (CSIC),
  the Ministerio de Econom\'ia y Competitividad (MINECO) and the European Regional Development Fund (ERDF) through projects FICTS-2011-02, ICTS-2017-07-CAHA-4, and CAHA16-CE-3978, 
  and the members of the CARMENES Consortium 
  (Max-Planck-Institut f\"ur Astronomie,
  Instituto de Astrof\'{\i}sica de Andaluc\'{\i}a,
  Landessternwarte K\"onigstuhl,
  Institut de Ci\`encies de l'Espai,
  Institut f\"ur Astrophysik G\"ottingen,
  Universidad Complutense de Madrid,
  Th\"uringer Landessternwarte Tautenburg,
  Instituto de Astrof\'{\i}sica de Canarias,
  Hamburger Sternwarte,
  Centro de Astrobiolog\'{\i}a and
  Centro Astron\'omico Hispano-Alem\'an), 
  with additional contributions by the MINECO, 
  the Deutsche Forschungsgemeinschaft through the Major Research Instrumentation Programme and Research Unit FOR2544 ``Blue Planets around Red Stars'', 
  the Klaus Tschira Stiftung, 
  the states of Baden-W\"urttemberg and Niedersachsen, 
  and by the Junta de Andaluc\'{\i}a.
  
  We acknowledge financial support from the Agencia Estatal de Investigaci\'on (AEI/10.13039/501100011033) of the Ministerio de Ciencia e Innovaci\'on and the ERDF ``A way of making Europe'' through projects 
  PID2022-137241NB-C4[1:4],	
  PID2021-125627OB-C31,		
  PID2019-109522GB-C5[1:4],	
and the Centre of Excellence ``Severo Ochoa'' and ``Mar\'ia de Maeztu'' awards to the Instituto de Astrof\'isica de Canarias (CEX2019-000920-S), Instituto de Astrof\'isica de Andaluc\'ia (SEV-2017-0709) and Institut de Ci\`encies de l'Espai (CEX2020-001058-M). 
We acknowledge support from the “Tecnologías avanzadas para la exploración de universo y sus componentes" (PR47/21 TAU) project funded by Comunidad de Madrid, by the Recovery, Transformation and Resilience Plan from the Spanish State, and by NextGenerationEU from the European Union through the Recovery and Resilience Facility.

This research has made use of the SIMBAD database, operated at CDS, Strasbourg, France, and the VALD database, operated at Uppsala University, the Institute of Astronomy RAS in Moscow, and the University of Vienna. 
 
\end{acknowledgements}

 \bibliographystyle{aa} % style aa.bst
\bibliography{MgSi} % your references Yourfile.bib

\begin{thebibliography}{99}
\expandafter\ifx\csname natexlab\endcsname\relax\def\natexlab#1{#1}\fi

\bibitem[{{Abia} {et~al.}(2020){Abia}, {Tabernero}, {Korotin}, {Montes}, {Marfil}, {Caballero}, {Straniero}, {Prantzos}, {Ribas}, {Reiners}, {Quirrenbach}, {Amado}, {B{\'e}jar}, {Cort{\'e}s-Contreras}, {Dreizler}, {Henning}, {Jeffers}, {Kaminski}, {K{\"u}rster}, {Lafarga}, {L{\'o}pez-Gallifa}, {Morales}, {Nagel}, {Passegger}, {Pedraz}, {Rodr{\'\i}guez L{\'o}pez}, {Schweitzer}, \& {Zechmeister}}]{abi20}
{Abia}, C., {Tabernero}, H.~M., {Korotin}, S.~A., {et~al.} 2020, \aap, 642, A227

\bibitem[{{Adams} \& {Laughlin}(1997)}]{ada97}
{Adams}, F.~C. \& {Laughlin}, G. 1997, Reviews of Modern Physics, 69, 337

\bibitem[{{Adibekyan} {et~al.}(2021){Adibekyan}, {Dorn}, {Sousa}, {Santos}, {Bitsch}, {Israelian}, {Mordasini}, {Barros}, {Delgado Mena}, {Demangeon}, {Faria}, {Figueira}, {Hakobyan}, {Oshagh}, {Soares}, {Kunitomo}, {Takeda}, {Jofr{\'e}}, {Petrucci}, \& {Martioli}}]{adi21}
{Adibekyan}, V., {Dorn}, C., {Sousa}, S.~G., {et~al.} 2021, Science, 374, 330

\bibitem[{{Adibekyan} {et~al.}(2015){Adibekyan}, {Santos}, {Figueira}, {Dorn}, {Sousa}, {Delgado-Mena}, {Israelian}, {Hakobyan}, \& {Mordasini}}]{adi15}
{Adibekyan}, V., {Santos}, N.~C., {Figueira}, P., {et~al.} 2015, \aap, 581, L2

\bibitem[{{Adibekyan} {et~al.}(2012){Adibekyan}, {Sousa}, {Santos}, {Delgado Mena}, {Gonz{\'a}lez Hern{\'a}ndez}, {Israelian}, {Mayor}, \& {Khachatryan}}]{adi12}
{Adibekyan}, V.~Z., {Sousa}, S.~G., {Santos}, N.~C., {et~al.} 2012, \aap, 545, A32

\bibitem[{{Allard} {et~al.}(2001){Allard}, {Hauschildt}, {Alexander}, {Tamanai}, \& {Schweitzer}}]{all01}
{Allard}, F., {Hauschildt}, P.~H., {Alexander}, D.~R., {Tamanai}, A., \& {Schweitzer}, A. 2001, \apj, 556, 357

\bibitem[{{Allard} {et~al.}(2012){Allard}, {Homeier}, \& {Freytag}}]{all12}
{Allard}, F., {Homeier}, D., \& {Freytag}, B. 2012, Philosophical Transactions of the Royal Society of London Series A, 370, 2765

\bibitem[{{Asplund} {et~al.}(2009){Asplund}, {Grevesse}, {Sauval}, \& {Scott}}]{asp09}
{Asplund}, M., {Grevesse}, N., {Sauval}, A.~J., \& {Scott}, P. 2009, \araa, 47, 481

\bibitem[{{Barber} {et~al.}(2006){Barber}, {Tennyson}, {Harris}, \& {Tolchenov}}]{Bar06}
{Barber}, R.~J., {Tennyson}, J., {Harris}, G.~J., \& {Tolchenov}, R.~N. 2006, \mnras, 368, 1087

\bibitem[{{Bensby} {et~al.}(2011){Bensby}, {Ad{\'e}n}, {Mel{\'e}ndez}, {Gould}, {Feltzing}, {Asplund}, {Johnson}, {Lucatello}, {Yee}, {Ram{\'\i}rez}, {Cohen}, {Thompson}, {Bond}, {Gal-Yam}, {Han}, {Sumi}, {Suzuki}, {Wada}, {Miyake}, {Furusawa}, {Ohmori}, {Saito}, {Tristram}, \& {Bennett}}]{ben11}
{Bensby}, T., {Ad{\'e}n}, D., {Mel{\'e}ndez}, J., {et~al.} 2011, \aap, 533, A134

\bibitem[{{Bensby} {et~al.}(2003){Bensby}, {Feltzing}, \& {Lundstr{\"o}m}}]{ben03}
{Bensby}, T., {Feltzing}, S., \& {Lundstr{\"o}m}, I. 2003, \aap, 410, 527

\bibitem[{{Bensby} {et~al.}(2005){Bensby}, {Feltzing}, {Lundstr{\"o}m}, \& {Ilyin}}]{ben05}
{Bensby}, T., {Feltzing}, S., {Lundstr{\"o}m}, I., \& {Ilyin}, I. 2005, \aap, 433, 185

\bibitem[{{Bitsch} \& {Battistini}(2020)}]{bit20}
{Bitsch}, B. \& {Battistini}, C. 2020, \aap, 633, A10

\bibitem[{{Bourrier} {et~al.}(2018){Bourrier}, {Dumusque}, {Dorn}, {Henry}, {Astudillo-Defru}, {Rey}, {Benneke}, {H{\'e}brard}, {Lovis}, {Demory}, {Moutou}, \& {Ehrenreich}}]{bou18}
{Bourrier}, V., {Dumusque}, X., {Dorn}, C., {et~al.} 2018, \aap, 619, A1

\bibitem[{{Brewer} \& {Fischer}(2016)}]{brew16b}
{Brewer}, J.~M. \& {Fischer}, D.~A. 2016, \apj, 831, 20

\bibitem[{{Brewer} {et~al.}(2016){Brewer}, {Fischer}, {Valenti}, \& {Piskunov}}]{brew16}
{Brewer}, J.~M., {Fischer}, D.~A., {Valenti}, J.~A., \& {Piskunov}, N. 2016, \apjs, 225, 32

\bibitem[{{Buchhave} {et~al.}(2018){Buchhave}, {Bitsch}, {Johansen}, {Latham}, {Bizzarro}, {Bieryla}, \& {Kipping}}]{buc18}
{Buchhave}, L.~A., {Bitsch}, B., {Johansen}, A., {et~al.} 2018, \apj, 856, 37

\bibitem[{{Buder} {et~al.}(2021){Buder}, {Sharma}, {Kos}, {Amarsi}, {Nordlander}, {Lind}, {Martell}, {Asplund}, {Bland-Hawthorn}, {Casey}, {de Silva}, {D'Orazi}, {Freeman}, {Hayden}, {Lewis}, {Lin}, {Schlesinger}, {Simpson}, {Stello}, {Zucker}, {Zwitter}, {Beeson}, {Buck}, {Casagrande}, {Clark}, {{\v{C}}otar}, {da Costa}, {de Grijs}, {Feuillet}, {Horner}, {Kafle}, {Khanna}, {Kobayashi}, {Liu}, {Montet}, {Nandakumar}, {Nataf}, {Ness}, {Spina}, {Tepper-Garc{\'\i}a}, {Ting}, {Traven}, {Vogrin{\v{c}}i{\v{c}}}, {Wittenmyer}, {Wyse}, {{\v{Z}}erjal}, \& {Galah Collaboration}}]{bud21}
{Buder}, S., {Sharma}, S., {Kos}, J., {et~al.} 2021, \mnras, 506, 150

\bibitem[{{Caballero} {et~al.}(2022){Caballero}, {Gonz{\'a}lez-{\'A}lvarez}, {Brady}, {Trifonov}, {Ellis}, {Dorn}, {Cifuentes}, {Molaverdikhani}, {Bean}, {Boyajian}, {Rodr{\'\i}guez}, {Sanz-Forcada}, {Zapatero Osorio}, {Abia}, {Amado}, {Anugu}, {B{\'e}jar}, {Davies}, {Dreizler}, {Dubois}, {Ennis}, {Espinoza}, {Farrington}, {L{\'o}pez}, {Gardner}, {Hatzes}, {Henning}, {Herrero}, {Herrero-Cisneros}, {Kaminski}, {Kasper}, {Klement}, {Kraus}, {Labdon}, {Lanthermann}, {Le Bouquin}, {L{\'o}pez Gonz{\'a}lez}, {Luque}, {Mann}, {Marfil}, {Monnier}, {Montes}, {Morales}, {Pall{\'e}}, {Pedraz}, {Quirrenbach}, {Reffert}, {Reiners}, {Ribas}, {Rodr{\'\i}guez-L{\'o}pez}, {Schaefer}, {Schweitzer}, {Seifahrt}, {Setterholm}, {Shan}, {Shulyak}, {Solano}, {Sreenivas}, {Stef{\'a}nsson}, {St{\"u}rmer}, {Tabernero}, {Tal-Or}, {ten Brummelaar}, {Vanaverbeke}, {von Braun}, {Youngblood}, \& {Zechmeister}}]{cab22}
{Caballero}, J.~A., {Gonz{\'a}lez-{\'A}lvarez}, E., {Brady}, M., {et~al.} 2022, \aap, 665, A120

\bibitem[{{Caballero} {et~al.}(2016){Caballero}, {Gu{\`a}rdia}, {L{\'o}pez del Fresno}, {Zechmeister}, {de Juan}, {Alonso-Floriano}, {Amado}, {Colom{\'e}}, {Cort{\'e}s-Contreras}, {Garc{\'\i}a-Piquer}, {Gesa}, {de Guindos}, {Hagen}, {Helmling}, {Hern{\'a}ndez Casta{\~n}o}, {K{\"u}rster}, {L{\'o}pez-Santiago}, {Montes}, {Morales Mu{\~n}oz}, {Pavlov}, {Quirrenbach}, {Reiners}, {Ribas}, {Seifert}, \& {Solano}}]{Cab16b}
{Caballero}, J.~A., {Gu{\`a}rdia}, J., {L{\'o}pez del Fresno}, M., {et~al.} 2016, in Society of Photo-Optical Instrumentation Engineers (SPIE) Conference Series, Vol. 9910, Observatory Operations: Strategies, Processes, and Systems VI, 99100E

\bibitem[{{Cifuentes} {et~al.}(2020){Cifuentes}, {Caballero}, {Cort{\'e}s-Contreras}, {Montes}, {Abell{\'a}n}, {Dorda}, {Holgado}, {Zapatero Osorio}, {Morales}, {Amado}, {Passegger}, {Quirrenbach}, {Reiners}, {Ribas}, {Sanz-Forcada}, {Schweitzer}, {Seifert}, \& {Solano}}]{cif20}
{Cifuentes}, C., {Caballero}, J.~A., {Cort{\'e}s-Contreras}, M., {et~al.} 2020, \aap, 642, A115

\bibitem[{{De Silva} {et~al.}(2015){De Silva}, {Freeman}, {Bland-Hawthorn}, {Martell}, {de Boer}, {Asplund}, {Keller}, {Sharma}, {Zucker}, {Zwitter}, {Anguiano}, {Bacigalupo}, {Bayliss}, {Beavis}, {Bergemann}, {Campbell}, {Cannon}, {Carollo}, {Casagrande}, {Casey}, {Da Costa}, {D'Orazi}, {Dotter}, {Duong}, {Heger}, {Ireland}, {Kafle}, {Kos}, {Lattanzio}, {Lewis}, {Lin}, {Lind}, {Munari}, {Nataf}, {O'Toole}, {Parker}, {Reid}, {Schlesinger}, {Sheinis}, {Simpson}, {Stello}, {Ting}, {Traven}, {Watson}, {Wittenmyer}, {Yong}, \& {{\v Z}erjal}}]{sil15}
{De Silva}, G.~M., {Freeman}, K.~C., {Bland-Hawthorn}, J., {et~al.} 2015, \mnras, 449, 2604

\bibitem[{{Deal} {et~al.}(2018){Deal}, {Alecian}, {Lebreton}, {Goupil}, {Marques}, {LeBlanc}, {Morel}, \& {Pichon}}]{dea18}
{Deal}, M., {Alecian}, G., {Lebreton}, Y., {et~al.} 2018, \aap, 618, A10

\bibitem[{{Delgado Mena} {et~al.}(2010){Delgado Mena}, {Israelian}, {Gonz{\'a}lez Hern{\'a}ndez}, {Bond}, {Santos}, {Udry}, \& {Mayor}}]{del10}
{Delgado Mena}, E., {Israelian}, G., {Gonz{\'a}lez Hern{\'a}ndez}, J.~I., {et~al.} 2010, \apj, 725, 2349

\bibitem[{{Dietrich} {et~al.}(2023){Dietrich}, {Apai}, {Schlecker}, {Hardegree-Ullman}, {Rackham}, {Kurtovic}, {Molaverdikhani}, {Gabor}, {Henning}, {Chen}, {Mancini}, {Bixel}, {Gibbs}, {Boyle}, {Brown-Sevilla}, {Burn}, {Delage}, {Flores-Rivera}, {Franceschi}, {Pichierri}, {Savvidou}, {Syed}, {Bruni}, {Ip}, {Ngeow}, {Tsai}, {Lin}, {Hou}, {Hsiao}, {Lin}, {Lin}, {Basant}, \& {EDEN Project}}]{die23}
{Dietrich}, J., {Apai}, D., {Schlecker}, M., {et~al.} 2023, \aj, 165, 149

\bibitem[{{Dorn} {et~al.}(2017){Dorn}, {Hinkel}, \& {Venturini}}]{dorn17}
{Dorn}, C., {Hinkel}, N.~R., \& {Venturini}, J. 2017, \aap, 597, A38

\bibitem[{{Dulick} {et~al.}(2003){Dulick}, {Bauschlicher}, {Burrows}, {Sharp}, {Ram}, \& {Bernath}}]{Dul03}
{Dulick}, M., {Bauschlicher}, C.~W., J., {Burrows}, A., {et~al.} 2003, \apj, 594, 651

\bibitem[{{Fischer} \& {Valenti}(2005)}]{fisval05}
{Fischer}, D.~A. \& {Valenti}, J. 2005, \apj, 622, 1102

\bibitem[{{Gaia Collaboration} {et~al.}(2021){Gaia Collaboration}, {Brown}, {Vallenari}, {Prusti}, {de Bruijne}, {Babusiaux}, {Biermann}, {Creevey}, {Evans}, {Eyer}, {Hutton}, {Jansen}, {Jordi}, {Klioner}, {Lammers}, {Lindegren}, {Luri}, {Mignard}, {Panem}, {Pourbaix}, {Randich}, {Sartoretti}, {Soubiran}, {Walton}, {Arenou}, {Bailer-Jones}, {Bastian}, {Cropper}, {Drimmel}, {Katz}, {Lattanzi}, {van Leeuwen}, {Bakker}, {Cacciari}, {Casta{\~n}eda}, {De Angeli}, {Ducourant}, {Fabricius}, {Fouesneau}, {Fr{\'e}mat}, {Guerra}, {Guerrier}, {Guiraud}, {Jean-Antoine Piccolo}, {Masana}, {Messineo}, {Mowlavi}, {Nicolas}, {Nienartowicz}, {Pailler}, {Panuzzo}, {Riclet}, {Roux}, {Seabroke}, {Sordo}, {Tanga}, {Th{\'e}venin}, {Gracia-Abril}, {Portell}, {Teyssier}, {Altmann}, {Andrae}, {Bellas-Velidis}, {Benson}, {Berthier}, {Blomme}, {Brugaletta}, {Burgess}, {Busso}, {Carry}, {Cellino}, {Cheek}, {Clementini}, {Damerdji}, {Davidson}, {Delchambre}, {Dell'Oro}, {Fern{\'a}ndez-Hern{\'a}ndez}, {Galluccio}, {Garc{\'\i}a-Lario},
  {Garcia-Reinaldos}, {Gonz{\'a}lez-N{\'u}{\~n}ez}, {Gosset}, {Haigron}, {Halbwachs}, {Hambly}, {Harrison}, {Hatzidimitriou}, {Heiter}, {Hern{\'a}ndez}, {Hestroffer}, {Hodgkin}, {Holl}, {Jan{\ss}en}, {Jevardat de Fombelle}, {Jordan}, {Krone-Martins}, {Lanzafame}, {L{\"o}ffler}, {Lorca}, {Manteiga}, {Marchal}, {Marrese}, {Moitinho}, {Mora}, {Muinonen}, {Osborne}, {Pancino}, {Pauwels}, {Petit}, {Recio-Blanco}, {Richards}, {Riello}, {Rimoldini}, {Robin}, {Roegiers}, {Rybizki}, {Sarro}, {Siopis}, {Smith}, {Sozzetti}, {Ulla}, {Utrilla}, {van Leeuwen}, {van Reeven}, {Abbas}, {Abreu Aramburu}, {Accart}, {Aerts}, {Aguado}, {Ajaj}, {Altavilla}, {{\'A}lvarez}, {{\'A}lvarez Cid-Fuentes}, {Alves}, {Anderson}, {Anglada Varela}, {Antoja}, {Audard}, {Baines}, {Baker}, {Balaguer-N{\'u}{\~n}ez}, {Balbinot}, {Balog}, {Barache}, {Barbato}, {Barros}, {Barstow}, {Bartolom{\'e}}, {Bassilana}, {Bauchet}, {Baudesson-Stella}, {Becciani}, {Bellazzini}, {Bernet}, {Bertone}, {Bianchi}, {Blanco-Cuaresma}, {Boch}, {Bombrun}, {Bossini},
  {Bouquillon}, {Bragaglia}, {Bramante}, {Breedt}, {Bressan}, {Brouillet}, {Bucciarelli}, {Burlacu}, {Busonero}, {Butkevich}, {Buzzi}, {Caffau}, {Cancelliere}, {C{\'a}novas}, {Cantat-Gaudin}, {Carballo}, {Carlucci}, {Carnerero}, {Carrasco}, {Casamiquela}, {Castellani}, {Castro-Ginard}, {Castro Sampol}, {Chaoul}, {Charlot}, {Chemin}, {Chiavassa}, {Cioni}, {Comoretto}, {Cooper}, {Cornez}, {Cowell}, {Crifo}, {Crosta}, {Crowley}, {Dafonte}, {Dapergolas}, {David}, {David}, {de Laverny}, {De Luise}, {De March}, {De Ridder}, {de Souza}, {de Teodoro}, {de Torres}, {del Peloso}, {del Pozo}, {Delbo}, {Delgado}, {Delgado}, {Delisle}, {Di Matteo}, {Diakite}, {Diener}, {Distefano}, {Dolding}, {Eappachen}, {Edvardsson}, {Enke}, {Esquej}, {Fabre}, {Fabrizio}, {Faigler}, {Fedorets}, {Fernique}, {Fienga}, {Figueras}, {Fouron}, {Fragkoudi}, {Fraile}, {Franke}, {Gai}, {Garabato}, {Garcia-Gutierrez}, {Garc{\'\i}a-Torres}, {Garofalo}, {Gavras}, {Gerlach}, {Geyer}, {Giacobbe}, {Gilmore}, {Girona}, {Giuffrida}, {Gomel}, {Gomez},
  {Gonzalez-Santamaria}, {Gonz{\'a}lez-Vidal}, {Granvik}, {Guti{\'e}rrez-S{\'a}nchez}, {Guy}, {Hauser}, {Haywood}, {Helmi}, {Hidalgo}, {Hilger}, {H{\l}adczuk}, {Hobbs}, {Holland}, {Huckle}, {Jasniewicz}, {Jonker}, {Juaristi Campillo}, {Julbe}, {Karbevska}, {Kervella}, {Khanna}, {Kochoska}, {Kontizas}, {Kordopatis}, {Korn}, {Kostrzewa-Rutkowska}, {Kruszy{\'n}ska}, {Lambert}, {Lanza}, {Lasne}, {Le Campion}, {Le Fustec}, {Lebreton}, {Lebzelter}, {Leccia}, {Leclerc}, {Lecoeur-Taibi}, {Liao}, {Licata}, {Lindstr{\o}m}, {Lister}, {Livanou}, {Lobel}, {Madrero Pardo}, {Managau}, {Mann}, {Marchant}, {Marconi}, {Marcos Santos}, {Marinoni}, {Marocco}, {Marshall}, {Martin Polo}, {Mart{\'\i}n-Fleitas}, {Masip}, {Massari}, {Mastrobuono-Battisti}, {Mazeh}, {McMillan}, {Messina}, {Michalik}, {Millar}, {Mints}, {Molina}, {Molinaro}, {Moln{\'a}r}, {Montegriffo}, {Mor}, {Morbidelli}, {Morel}, {Morris}, {Mulone}, {Munoz}, {Muraveva}, {Murphy}, {Musella}, {Noval}, {Ord{\'e}novic}, {Orr{\`u}}, {Osinde}, {Pagani}, {Pagano},
  {Palaversa}, {Palicio}, {Panahi}, {Pawlak}, {Pe{\~n}alosa Esteller}, {Penttil{\"a}}, {Piersimoni}, {Pineau}, {Plachy}, {Plum}, {Poggio}, {Poretti}, {Poujoulet}, {Pr{\v{s}}a}, {Pulone}, {Racero}, {Ragaini}, {Rainer}, {Raiteri}, {Rambaux}, {Ramos}, {Ramos-Lerate}, {Re Fiorentin}, {Regibo}, {Reyl{\'e}}, {Ripepi}, {Riva}, {Rixon}, {Robichon}, {Robin}, {Roelens}, {Rohrbasser}, {Romero-G{\'o}mez}, {Rowell}, {Royer}, {Rybicki}, {Sadowski}, {Sagrist{\`a} Sell{\'e}s}, {Sahlmann}, {Salgado}, {Salguero}, {Samaras}, {Sanchez Gimenez}, {Sanna}, {Santove{\~n}a}, {Sarasso}, {Schultheis}, {Sciacca}, {Segol}, {Segovia}, {S{\'e}gransan}, {Semeux}, {Shahaf}, {Siddiqui}, {Siebert}, {Siltala}, {Slezak}, {Smart}, {Solano}, {Solitro}, {Souami}, {Souchay}, {Spagna}, {Spoto}, {Steele}, {Steidelm{\"u}ller}, {Stephenson}, {S{\"u}veges}, {Szabados}, {Szegedi-Elek}, {Taris}, {Tauran}, {Taylor}, {Teixeira}, {Thuillot}, {Tonello}, {Torra}, {Torra}, {Turon}, {Unger}, {Vaillant}, {van Dillen}, {Vanel}, {Vecchiato}, {Viala}, {Vicente},
  {Voutsinas}, {Weiler}, {Wevers}, {Wyrzykowski}, {Yoldas}, {Yvard}, {Zhao}, {Zorec}, {Zucker}, {Zurbach}, \& {Zwitter}}]{EDR3}
{Gaia Collaboration}, {Brown}, A.~G.~A., {Vallenari}, A., {et~al.} 2021, \aap, 649, A1

\bibitem[{{Ghezzi} {et~al.}(2010){Ghezzi}, {Cunha}, {Smith}, {de Ara{\'u}jo}, {Schuler}, \& {de la Reza}}]{ghe10}
{Ghezzi}, L., {Cunha}, K., {Smith}, V.~V., {et~al.} 2010, \apj, 720, 1290

\bibitem[{{Gilmore} {et~al.}(2022){Gilmore}, {Randich}, {Worley}, {Hourihane}, {Gonneau}, {Sacco}, {Lewis}, {Magrini}, {Fran{\c{c}}ois}, {Jeffries}, {Koposov}, {Bragaglia}, {Alfaro}, {Allende Prieto}, {Blomme}, {Korn}, {Lanzafame}, {Pancino}, {Recio-Blanco}, {Smiljanic}, {Van Eck}, {Zwitter}, {Bensby}, {Flaccomio}, {Irwin}, {Franciosini}, {Morbidelli}, {Damiani}, {Bonito}, {Friel}, {Vink}, {Prisinzano}, {Abbas}, {Hatzidimitriou}, {Held}, {Jordi}, {Paunzen}, {Spagna}, {Jackson}, {Ma{\'\i}z Apell{\'a}niz}, {Asplund}, {Bonifacio}, {Feltzing}, {Binney}, {Drew}, {Ferguson}, {Micela}, {Negueruela}, {Prusti}, {Rix}, {Vallenari}, {Bergemann}, {Casey}, {de Laverny}, {Frasca}, {Hill}, {Lind}, {Sbordone}, {Sousa}, {Adibekyan}, {Caffau}, {Daflon}, {Feuillet}, {Gebran}, {Gonzalez Hernandez}, {Guiglion}, {Herrero}, {Lobel}, {Merle}, {Mikolaitis}, {Montes}, {Morel}, {Ruchti}, {Soubiran}, {Tabernero}, {Tautvai{\v{s}}ien{\.{e}}}, {Traven}, {Valentini}, {Van der Swaelmen}, {Villanova}, {Viscasillas V{\'a}zquez}, {Bayo},
  {Biazzo}, {Carraro}, {Edvardsson}, {Heiter}, {Jofr{\'e}}, {Marconi}, {Martayan}, {Masseron}, {Monaco}, {Walton}, {Zaggia}, {Aguirre B{\o}rsen-Koch}, {Alves}, {Balaguer-Nunez}, {Barklem}, {Barrado}, {Bellazzini}, {Berlanas}, {Binks}, {Bressan}, {Capuzzo-Dolcetta}, {Casagrande}, {Casamiquela}, {Collins}, {D'Orazi}, {Dantas}, {Debattista}, {Delgado-Mena}, {Di Marcantonio}, {Drazdauskas}, {Evans}, {Famaey}, {Franchini}, {Fr{\'e}mat}, {Fu}, {Geisler}, {Gerhard}, {Gonz{\'a}lez Solares}, {Grebel}, {Guti{\'e}rrez Albarr{\'a}n}, {Jim{\'e}nez-Esteban}, {J{\"o}nsson}, {Khachaturyants}, {Kordopatis}, {Kos}, {Lagarde}, {Ludwig}, {Mahy}, {Mapelli}, {Marfil}, {Martell}, {Messina}, {Miglio}, {Minchev}, {Moitinho}, {Montalban}, {Monteiro}, {Morossi}, {Mowlavi}, {Mucciarelli}, {Murphy}, {Nardetto}, {Ortolani}, {Paletou}, {Palou{\v{s}}}, {Pickering}, {Quirrenbach}, {Re Fiorentin}, {Read}, {Romano}, {Ryde}, {Sanna}, {Santos}, {Seabroke}, {Spina}, {Steinmetz}, {Stonkut{\'e}}, {Sutorius}, {Th{\'e}venin}, {Tosi}, {Tsantaki},
  {Wright}, {Wyse}, {Zoccali}, {Zorec}, \& {Zucker}}]{gil22}
{Gilmore}, G., {Randich}, S., {Worley}, C.~C., {et~al.} 2022, \aap, 666, A120

\bibitem[{{Gonzalez}(1997)}]{gon97}
{Gonzalez}, G. 1997, \mnras, 285, 403

\bibitem[{{Gonzalez}(2006)}]{gon06}
{Gonzalez}, G. 2006, \pasp, 118, 1494

\bibitem[{{Gonz{\'a}lez Hern{\'a}ndez} {et~al.}(2013){Gonz{\'a}lez Hern{\'a}ndez}, {Delgado-Mena}, {Sousa}, {Israelian}, {Santos}, {Adibekyan}, \& {Udry}}]{gon13}
{Gonz{\'a}lez Hern{\'a}ndez}, J.~I., {Delgado-Mena}, E., {Sousa}, S.~G., {et~al.} 2013, \aap, 552, A6

\bibitem[{{Gonz{\'a}lez Hern{\'a}ndez} {et~al.}(2010){Gonz{\'a}lez Hern{\'a}ndez}, {Israelian}, {Santos}, {Sousa}, {Delgado-Mena}, {Neves}, \& {Udry}}]{gon10}
{Gonz{\'a}lez Hern{\'a}ndez}, J.~I., {Israelian}, G., {Santos}, N.~C., {et~al.} 2010, \apj, 720, 1592

\bibitem[{{Goorvitch}(1994)}]{Goo94}
{Goorvitch}, D. 1994, \apjs, 95, 535

\bibitem[{{Guillot} {et~al.}(2006){Guillot}, {Santos}, {Pont}, {Iro}, {Melo}, \& {Ribas}}]{gui06}
{Guillot}, T., {Santos}, N.~C., {Pont}, F., {et~al.} 2006, \aap, 453, L21

\bibitem[{{Heiter} {et~al.}(2021){Heiter}, {Lind}, {Bergemann}, {Asplund}, {Mikolaitis}, {Barklem}, {Masseron}, {de Laverny}, {Magrini}, {Edvardsson}, {J{\"o}nsson}, {Pickering}, {Ryde}, {Bayo Ar{\'a}n}, {Bensby}, {Casey}, {Feltzing}, {Jofr{\'e}}, {Korn}, {Pancino}, {Damiani}, {Lanzafame}, {Lardo}, {Monaco}, {Morbidelli}, {Smiljanic}, {Worley}, {Zaggia}, {Randich}, \& {Gilmore}}]{Hei21}
{Heiter}, U., {Lind}, K., {Bergemann}, M., {et~al.} 2021, \aap, 645, A106

\bibitem[{{Hejazi} {et~al.}(2023){Hejazi}, {Crossfield}, {Nordlander}, {Mansfield}, {Souto}, {Marfil}, {Coria}, {Brande}, {Polanski}, {Hand}, \& {Wienke}}]{hej23}
{Hejazi}, N., {Crossfield}, I. J.~M., {Nordlander}, T., {et~al.} 2023, \apj, 949, 79

\bibitem[{{Henry} {et~al.}(2006){Henry}, {Jao}, {Subasavage}, {Beaulieu}, {Ianna}, {Costa}, \& {M{\'e}ndez}}]{hen06}
{Henry}, T.~J., {Jao}, W.-C., {Subasavage}, J.~P., {et~al.} 2006, \aj, 132, 2360

\bibitem[{{Hogg} {et~al.}(2016){Hogg}, {Casey}, {Ness}, {Rix}, {Foreman-Mackey}, {Hasselquist}, {Ho}, {Holtzman}, {Majewski}, {Martell}, {M{\'e}sz{\'a}ros}, {Nidever}, \& {Shetrone}}]{hog16}
{Hogg}, D.~W., {Casey}, A.~R., {Ness}, M., {et~al.} 2016, \apj, 833, 262

\bibitem[{{Ishikawa} {et~al.}(2022){Ishikawa}, {Aoki}, {Hirano}, {Kotani}, {Kuzuhara}, {Omiya}, {Hori}, {Kokubo}, {Kudo}, {Kurokawa}, {Kusakabe}, {Narita}, {Nishikawa}, {Ogihara}, {Ueda}, {Currie}, {Henning}, {Kasagi}, {Kolecki}, {Kwon}, {Machida}, {McElwain}, {Nakagawa}, {Vievard}, {Wang}, {Tamura}, \& {Sato}}]{ish22}
{Ishikawa}, H.~T., {Aoki}, W., {Hirano}, T., {et~al.} 2022, \aj, 163, 72

\bibitem[{{Ishikawa} {et~al.}(2020){Ishikawa}, {Aoki}, {Kotani}, {Kuzuhara}, {Omiya}, {Reiners}, \& {Zechmeister}}]{ish20}
{Ishikawa}, H.~T., {Aoki}, W., {Kotani}, T., {et~al.} 2020, \pasj, 72, 102

\bibitem[{{Jahandar} {et~al.}(2024){Jahandar}, {Doyon}, {Artigau}, {Cook}, {Cadieux}, {Lafreni{\`e}re}, {Forveille}, {Donati}, {Fouqu{\'e}}, {Carmona}, {Cloutier}, {Cristofari}, {Gaidos}, {Gomes da Silva}, {Malo}, {Martioli}, {do Nascimento}, {Pelletier}, {Vandal}, \& {Venn}}]{jah23}
{Jahandar}, F., {Doyon}, R., {Artigau}, {\'E}., {et~al.} 2024, \apj, 966, 56

\bibitem[{{Johnson} \& {Soderblom}(1987)}]{uvw87}
{Johnson}, D. R.~H. \& {Soderblom}, D.~R. 1987, \aj, 93, 864

\bibitem[{{Johnson} {et~al.}(2010){Johnson}, {Aller}, {Howard}, \& {Crepp}}]{joh10}
{Johnson}, J.~A., {Aller}, K.~M., {Howard}, A.~W., \& {Crepp}, J.~R. 2010, \pasp, 122, 905

\bibitem[{{Kausch} {et~al.}(2015){Kausch}, {Noll}, {Smette}, {Kimeswenger}, {Barden}, {Szyszka}, {Jones}, {Sana}, {Horst}, \& {Kerber}}]{kau15}
{Kausch}, W., {Noll}, S., {Smette}, A., {et~al.} 2015, \aap, 576, A78

\bibitem[{{Kirkpatrick} {et~al.}(1991){Kirkpatrick}, {Henry}, \& {McCarthy}}]{kir91}
{Kirkpatrick}, J.~D., {Henry}, T.~J., \& {McCarthy}, Donald~W., J. 1991, \apjs, 77, 417

\bibitem[{{Kobayashi} {et~al.}(2020){Kobayashi}, {Karakas}, \& {Lugaro}}]{kob20}
{Kobayashi}, C., {Karakas}, A.~I., \& {Lugaro}, M. 2020, \apj, 900, 179

\bibitem[{{Korn} {et~al.}(2007){Korn}, {Grundahl}, {Richard}, {Mashonkina}, {Barklem}, {Collet}, {Gustafsson}, \& {Piskunov}}]{kor07}
{Korn}, A.~J., {Grundahl}, F., {Richard}, O., {et~al.} 2007, \apj, 671, 402

\bibitem[{{Kurucz}(2014)}]{Kur14}
{Kurucz}, R.~L. 2014, {Problems with Atomic and Molecular Data: Including All the Lines} (Springer, Cham), 63--73

\bibitem[{{Lafarga} {et~al.}(2020){Lafarga}, {Ribas}, {Lovis}, {Perger}, {Zechmeister}, {Bauer}, {K{\"u}rster}, {Cort{\'e}s-Contreras}, {Morales}, {Herrero}, {Rosich}, {Baroch}, {Reiners}, {Caballero}, {Quirrenbach}, {Amado}, {Alacid}, {B{\'e}jar}, {Dreizler}, {Hatzes}, {Henning}, {Jeffers}, {Kaminski}, {Montes}, {Pedraz}, {Rodr{\'\i}guez-L{\'o}pez}, \& {Schmitt}}]{laf20}
{Lafarga}, M., {Ribas}, I., {Lovis}, C., {et~al.} 2020, \aap, 636, A36

\bibitem[{{Laughlin} {et~al.}(2004){Laughlin}, {Bodenheimer}, \& {Adams}}]{lau04}
{Laughlin}, G., {Bodenheimer}, P., \& {Adams}, F.~C. 2004, \apjl, 612, L73

\bibitem[{{Lichtenberg} {et~al.}(2021){Lichtenberg}, {Dr{\k{a}}{\.z}kowska}, {Sch{\"o}nb{\"a}chler}, {Golabek}, \& {Hands}}]{lich21}
{Lichtenberg}, T., {Dr{\k{a}}{\.z}kowska}, J., {Sch{\"o}nb{\"a}chler}, M., {Golabek}, G.~J., \& {Hands}, T.~O. 2021, Science, 371, 365

\bibitem[{{Lodders}(2003)}]{lod03}
{Lodders}, K. 2003, \apj, 591, 1220

\bibitem[{{Majewski} {et~al.}(2017){Majewski}, {Schiavon}, {Frinchaboy}, {Allende Prieto}, {Barkhouser}, {Bizyaev}, {Blank}, {Brunner}, {Burton}, {Carrera}, {Chojnowski}, {Cunha}, {Epstein}, {Fitzgerald}, {Garc{\'\i}a P{\'e}rez}, {Hearty}, {Henderson}, {Holtzman}, {Johnson}, {Lam}, {Lawler}, {Maseman}, {M{\'e}sz{\'a}ros}, {Nelson}, {Nguyen}, {Nidever}, {Pinsonneault}, {Shetrone}, {Smee}, {Smith}, {Stolberg}, {Skrutskie}, {Walker}, {Wilson}, {Zasowski}, {Anders}, {Basu}, {Beland}, {Blanton}, {Bovy}, {Brownstein}, {Carlberg}, {Chaplin}, {Chiappini}, {Eisenstein}, {Elsworth}, {Feuillet}, {Fleming}, {Galbraith-Frew}, {Garc{\'\i}a}, {Garc{\'\i}a-Hern{\'a}ndez}, {Gillespie}, {Girardi}, {Gunn}, {Hasselquist}, {Hayden}, {Hekker}, {Ivans}, {Kinemuchi}, {Klaene}, {Mahadevan}, {Mathur}, {Mosser}, {Muna}, {Munn}, {Nichol}, {O'Connell}, {Parejko}, {Robin}, {Rocha-Pinto}, {Schultheis}, {Serenelli}, {Shane}, {Silva Aguirre}, {Sobeck}, {Thompson}, {Troup}, {Weinberg}, \& {Zamora}}]{maj17}
{Majewski}, S.~R., {Schiavon}, R.~P., {Frinchaboy}, P.~M., {et~al.} 2017, \aj, 154, 94

\bibitem[{{Maldonado} {et~al.}(2020){Maldonado}, {Micela}, {Baratella}, {D'Orazi}, {Affer}, {Biazzo}, {Lanza}, {Maggio}, {Gonz{\'a}lez Hern{\'a}ndez}, {Perger}, {Pinamonti}, {Scandariato}, {Sozzetti}, {Locci}, {Di Maio}, {Bignamini}, {Claudi}, {Molinari}, {Rebolo}, {Ribas}, {Toledo-Padr{\'o}n}, {Covino}, {Desidera}, {Herrero}, {Morales}, {Su{\'a}rez-Mascare{\~n}o}, {Pagano}, {Petralia}, {Piotto}, \& {Poretti}}]{mal20}
{Maldonado}, J., {Micela}, G., {Baratella}, M., {et~al.} 2020, \aap, 644, A68

\bibitem[{{Marfil} {et~al.}(2021){Marfil}, {Tabernero}, {Montes}, {Caballero}, {L{\'a}zaro}, {Gonz{\'a}lez Hern{\'a}ndez}, {Nagel}, {Passegger}, {Schweitzer}, {Ribas}, {Reiners}, {Quirrenbach}, {Amado}, {Cifuentes}, {Cort{\'e}s-Contreras}, {Dreizler}, {Duque-Arribas}, {Galad{\'\i}-Enr{\'\i}quez}, {Henning}, {Jeffers}, {Kaminski}, {K{\"u}rster}, {Lafarga}, {L{\'o}pez-Gallifa}, {Morales}, {Shan}, \& {Zechmeister}}]{mar21}
{Marfil}, E., {Tabernero}, H.~M., {Montes}, D., {et~al.} 2021, \aap, 656, A162

\bibitem[{{McKemmish} {et~al.}(2016){McKemmish}, {Yurchenko}, \& {Tennyson}}]{kem16}
{McKemmish}, L.~K., {Yurchenko}, S.~N., \& {Tennyson}, J. 2016, \mnras, 463, 771

\bibitem[{{Mitschang} {et~al.}(2014){Mitschang}, {De Silva}, {Zucker}, {Anguiano}, {Bensby}, \& {Feltzing}}]{mit14}
{Mitschang}, A.~W., {De Silva}, G., {Zucker}, D.~B., {et~al.} 2014, \mnras, 438, 2753

\bibitem[{{Montes} {et~al.}(2018){Montes}, {Gonz{\'a}lez-Peinado}, {Tabernero}, {Caballero}, {Marfil}, {Alonso-Floriano}, {Cort{\'e}s-Contreras}, {Gonz{\'a}lez Hern{\'a}ndez}, {Klutsch}, \& {Moreno-J{\'o}dar}}]{mon18}
{Montes}, D., {Gonz{\'a}lez-Peinado}, R., {Tabernero}, H.~M., {et~al.} 2018, \mnras, 479, 1332

\bibitem[{{Montes} {et~al.}(2001){Montes}, {L{\'o}pez-Santiago}, {G{\'a}lvez}, {Fern{\'a}ndez-Figueroa}, {De Castro}, \& {Cornide}}]{mon01}
{Montes}, D., {L{\'o}pez-Santiago}, J., {G{\'a}lvez}, M.~C., {et~al.} 2001, \mnras, 328, 45

\bibitem[{{Morgan} \& {Anders}(1980)}]{morand80}
{Morgan}, J.~W. \& {Anders}, E. 1980, Proceedings of the National Academy of Science, 77, 6973

\bibitem[{{Morton}(2000)}]{mor00}
{Morton}, D.~C. 2000, \apjs, 130, 403

\bibitem[{{Nagel} {et~al.}(2023){Nagel}, {Czesla}, {Kaminski}, {Zechmeister}, {Tal-Or}, {Schmitt}, {Reiners}, {Quirrenbach}, {Garc{\'\i}a L{\'o}pez}, {Caballero}, {Ribas}, {Amado}, {B{\'e}jar}, {Cort{\'e}s-Contreras}, {Dreizler}, {Hatzes}, {Henning}, {Jeffers}, {K{\"u}rster}, {Lafarga}, {L{\'o}pez-Puertas}, {Montes}, {Morales}, {Pedraz}, \& {Schweitzer}}]{nag23}
{Nagel}, E., {Czesla}, S., {Kaminski}, A., {et~al.} 2023, \aap, 680, A73

\bibitem[{{Passegger} {et~al.}(2018){Passegger}, {Reiners}, {Jeffers}, {Wende-von Berg}, {Sch{\"o}fer}, {Caballero}, {Schweitzer}, {Amado}, {B{\'e}jar}, {Cort{\'e}s-Contreras}, {Hatzes}, {K{\"u}rster}, {Montes}, {Pedraz}, {Quirrenbach}, {Ribas}, \& {Seifert}}]{pas18}
{Passegger}, V.~M., {Reiners}, A., {Jeffers}, S.~V., {et~al.} 2018, \aap, 615, A6

\bibitem[{{Perger} {et~al.}(2019){Perger}, {Scandariato}, {Ribas}, {Morales}, {Affer}, {Azzaro}, {Amado}, {Anglada-Escud{\'e}}, {Baroch}, {Barrado}, {Bauer}, {B{\'e}jar}, {Caballero}, {Cort{\'e}s-Contreras}, {Damasso}, {Dreizler}, {Gonz{\'a}lez-Cuesta}, {Gonz{\'a}lez Hern{\'a}ndez}, {Guenther}, {Henning}, {Herrero}, {Jeffers}, {Kaminski}, {K{\"u}rster}, {Lafarga}, {Leto}, {L{\'o}pez-Gonz{\'a}lez}, {Maldonado}, {Micela}, {Montes}, {Pinamonti}, {Quirrenbach}, {Rebolo}, {Reiners}, {Rodr{\'\i}guez}, {Rodr{\'\i}guez-L{\'o}pez}, {Schmitt}, {Sozzetti}, {Su{\'a}rez Mascare{\~n}o}, {Toledo-Padr{\'o}n}, {Zanmar S{\'a}nchez}, {Zapatero Osorio}, \& {Zechmeister}}]{per19}
{Perger}, M., {Scandariato}, G., {Ribas}, I., {et~al.} 2019, \aap, 624, A123

\bibitem[{{Plez}(2012)}]{ple12}
{Plez}, B. 2012, {Turbospectrum: Code for spectral synthesis}, Astrophysics Source Code Library

\bibitem[{{Quirrenbach} {et~al.}(2014){Quirrenbach}, {Amado}, {Caballero}, {Mundt}, {Reiners}, {Ribas}, {Seifert}, {Abril}, {Aceituno}, {Alonso-Floriano}, {Ammler-von Eiff}, {Antona Jim{\'e}nez}, {Anwand-Heerwart}, {Azzaro}, {Bauer}, {Barrado}, {Becerril}, {B{\'e}jar}, {Ben{\'\i}tez}, {Berdi{\~n}as}, {C{\'a}rdenas}, {Casal}, {Claret}, {Colom{\'e}}, {Cort{\'e}s-Contreras}, {Czesla}, {Doellinger}, {Dreizler}, {Feiz}, {Fern{\'a}ndez}, {Galad{\'\i}}, {G{\'a}lvez-Ortiz}, {Garc{\'\i}a-Piquer}, {Garc{\'\i}a-Vargas}, {Garrido}, {Gesa}, {G{\'o}mez Galera}, {Gonz{\'a}lez {\'A}lvarez}, {Gonz{\'a}lez Hern{\'a}ndez}, {Gr{\"o}zinger}, {Gu{\`a}rdia}, {Guenther}, {de Guindos}, {Guti{\'e}rrez-Soto}, {Hagen}, {Hatzes}, {Hauschildt}, {Helmling}, {Henning}, {Hermann}, {Hern{\'a}ndez Casta{\~n}o}, {Herrero}, {Hidalgo}, {Holgado}, {Huber}, {Huber}, {Jeffers}, {Joergens}, {de Juan}, {Kehr}, {Klein}, {K{\"u}rster}, {Lamert}, {Lalitha}, {Laun}, {Lemke}, {Lenzen}, {L{\'o}pez del Fresno}, {L{\'o}pez Mart{\'\i}}, {L{\'o}pez-Santiago},
  {Mall}, {Mandel}, {Mart{\'\i}n}, {Mart{\'\i}n-Ruiz}, {Mart{\'\i}nez-Rodr{\'\i}guez}, {Marvin}, {Mathar}, {Mirabet}, {Montes}, {Morales Mu{\~n}oz}, {Moya}, {Naranjo}, {Ofir}, {Oreiro}, {Pall{\'e}}, {Panduro}, {Passegger}, {P{\'e}rez-Calpena}, {P{\'e}rez Medialdea}, {Perger}, {Pluto}, {Ram{\'o}n}, {Rebolo}, {Redondo}, {Reffert}, {Reinhardt}, {Rhode}, {Rix}, {Rodler}, {Rodr{\'\i}guez}, {Rodr{\'\i}guez-L{\'o}pez}, {Rodr{\'\i}guez-P{\'e}rez}, {Rohloff}, {Rosich}, {S{\'a}nchez-Blanco}, {S{\'a}nchez Carrasco}, {Sanz-Forcada}, {Sarmiento}, {Sch{\"a}fer}, {Schiller}, {Schmidt}, {Schmitt}, {Solano}, {Stahl}, {Storz}, {St{\"u}rmer}, {Su{\'a}rez}, {Ulbrich}, {Veredas}, {Wagner}, {Winkler}, {Zapatero Osorio}, {Zechmeister}, {Abell{\'a}n de Paco}, {Anglada-Escud{\'e}}, {del Burgo}, {Klutsch}, {Lizon}, {L{\'o}pez-Morales}, {Morales}, {Perryman}, {Tulloch}, \& {Xu}}]{quir14}
{Quirrenbach}, A., {Amado}, P.~J., {Caballero}, J.~A., {et~al.} 2014, in Society of Photo-Optical Instrumentation Engineers (SPIE) Conference Series, Vol. 9147, Ground-based and Airborne Instrumentation for Astronomy V, ed. S.~K. {Ramsay}, I.~S. {McLean}, \& H.~{Takami}, 91471F

\bibitem[{{Quirrenbach} {et~al.}(2020){Quirrenbach}, {CARMENES Consortium}, {Amado}, {Ribas}, {Reiners}, {Caballero}, {Aceituno}, {Alacid}, {Alonso-Floriano}, {Anglada-Escud{\'e}}, {Azzaro}, {Baroch}, {Bauer}, {Becerril}, {B{\'e}jar}, {Bluhm}, {Calvo Ortega}, {Cardona Guill{\'e}n}, {Casasayas-Barris}, {Chaturvedi}, {Cifuentes}, {Colom{\'e}}, {Conte}, {Cort{\'e}s-Contreras}, {Czesla}, {D{\'\i}ez-Alonso}, {Dom{\'\i}nguez Fern{\'a}ndez}, {Dreizler}, {Duque-Arribas}, {Espinoza}, {Fuhrmeister}, {Galad{\'\i}-Enr{\'\i}quez}, {Gar{\textasciiacute}a Quintana}, {Gonz{\'a}lez-Alvare}, {Gonz{\'a}lez Cuesta}, {Gonz{\'a}lez Hern{\'a}ndez}, {Guenther}, {de Guindos}, {Hatzes}, {Henning}, {Herbort}, {Herrero}, {Hintz}, {Iglesias-P{\'a}ra}, {Jeffers}, {Johnson}, {de Juan}, {Kaminski}, {Kemmer}, {Khaimova}, {Khalafinejad}, {Klahr}, {Kossakowski}, {Kreidberg}, {K{\"u}rster}, {Labarga}, {Lafarga}, {Lamp{\'o}n}, {Lara}, {Lillo-Box}, {Lodieu}, {L{\'o}pez Gallifa}, {L{\'o}pez Gonz{\'a}lez}, {L{\'o}pez-Puertas}, {Luque}, {Marfil},
  {Mart{\'\i}n-Ruiz}, {Matth{\'e}}, {Molaverdikhani}, {Montes}, {Morales}, {Morales-Calder{\'o}on}, {Nagel}, {Nortmann}, {Nowak}, {Ofir}, {Oshaghi}, {Pall{\'e}}, {Passegger}, {Pavlov}, {Pedraz}, {Perdelwitz}, {Perger}, {Reffert}, {Revilla}, {Rodr{\'\i}guez}, {Rodr{\'\i}guez L{\'o}pez}, {Sabotta}, {Sadegi}, {Sairam}, {Salz}, {S{\'a}nchez-L{\'o}pez}, {Sanz-Forcada}, {Sarkis}, {Sch{\"a}fer}, {Schiller}, {Schlecker}, {Schmitt}, {Sch{\"o}fer}, {Schweitzer}, {Seiferta}, {Shan}, {Shulyak}, {Skrzypinski}, {Solano}, {Soto}, {Stahl}, {Stangret}, {Stock}, {Strachan}, {Stuber}, {St{\"u}rmer}, {Tabernero}, {Tal-Or}, {Tala-Pinto}, {Trifonov}, {Vanaverbeke}, {Yan}, {Zapatero Osorio}, \& {Zechmeister}}]{quir20}
{Quirrenbach}, A., {CARMENES Consortium}, {Amado}, P.~J., {et~al.} 2020, in Society of Photo-Optical Instrumentation Engineers (SPIE) Conference Series, Vol. 11447, Society of Photo-Optical Instrumentation Engineers (SPIE) Conference Series, 114473C

\bibitem[{{Rajpurohit} {et~al.}(2018){Rajpurohit}, {Allard}, {Teixeira}, {Homeier}, {Rajpurohit}, \& {Mousis}}]{raj18}
{Rajpurohit}, A.~S., {Allard}, F., {Teixeira}, G.~D.~C., {et~al.} 2018, \aap, 610, A19

\bibitem[{{Randich} {et~al.}(2022){Randich}, {Gilmore}, {Magrini}, {Sacco}, {Jackson}, {Jeffries}, {Worley}, {Hourihane}, {Gonneau}, {Viscasillas Vazquez}, {Franciosini}, {Lewis}, {Alfaro}, {Allende Prieto}, {Bensby}, {Blomme}, {Bragaglia}, {Flaccomio}, {Fran{\c{c}}ois}, {Irwin}, {Koposov}, {Korn}, {Lanzafame}, {Pancino}, {Recio-Blanco}, {Smiljanic}, {Van Eck}, {Zwitter}, {Asplund}, {Bonifacio}, {Feltzing}, {Binney}, {Drew}, {Ferguson}, {Micela}, {Negueruela}, {Prusti}, {Rix}, {Vallenari}, {Bayo}, {Bergemann}, {Biazzo}, {Carraro}, {Casey}, {Damiani}, {Frasca}, {Heiter}, {Hill}, {Jofr{\'e}}, {de Laverny}, {Lind}, {Marconi}, {Martayan}, {Masseron}, {Monaco}, {Morbidelli}, {Prisinzano}, {Sbordone}, {Sousa}, {Zaggia}, {Adibekyan}, {Bonito}, {Caffau}, {Daflon}, {Feuillet}, {Gebran}, {Gonzalez Hernandez}, {Guiglion}, {Herrero}, {Lobel}, {Maiz Apellaniz}, {Merle}, {Mikolaitis}, {Montes}, {Morel}, {Soubiran}, {Spina}, {Tabernero}, {Tautvai{\v{s}}iene}, {Traven}, {Valentini}, {Van der Swaelmen}, {Villanova}, {Wright},
  {Abbas}, {Aguirre B{\o}rsen-Koch}, {Alves}, {Balaguer-Nunez}, {Barklem}, {Barrado}, {Berlanas}, {Binks}, {Bressan}, {Capuzzo-Dolcetta}, {Casagrande}, {Casamiquela}, {Collins}, {D'Orazi}, {Dantas}, {Debattista}, {Delgado-Mena}, {Di Marcantonio}, {Drazdauskas}, {Evans}, {Famaey}, {Franchini}, {Fr{\'e}mat}, {Friel}, {Fu}, {Geisler}, {Gerhard}, {Gonzalez Solares}, {Grebel}, {Gutierrez Albarran}, {Hatzidimitriou}, {Held}, {Jim{\'e}nez-Esteban}, {J{\"o}nsson}, {Jordi}, {Khachaturyants}, {Kordopatis}, {Kos}, {Lagarde}, {Mahy}, {Mapelli}, {Marfil}, {Martell}, {Messina}, {Miglio}, {Minchev}, {Moitinho}, {Montalban}, {Monteiro}, {Morossi}, {Mowlavi}, {Mucciarelli}, {Murphy}, {Nardetto}, {Ortolani}, {Paletou}, {Palou{\v{s}}}, {Paunzen}, {Pickering}, {Quirrenbach}, {Re Fiorentin}, {Read}, {Romano}, {Ryde}, {Sanna}, {Santos}, {Seabroke}, {Spagna}, {Steinmetz}, {Stonkut{\'e}}, {Sutorius}, {Th{\'e}venin}, {Tosi}, {Tsantaki}, {Vink}, {Wright}, {Wyse}, {Zoccali}, {Zorec}, {Zucker}, \& {Walton}}]{ran22}
{Randich}, S., {Gilmore}, G., {Magrini}, L., {et~al.} 2022, \aap, 666, A121

\bibitem[{{Reffert} {et~al.}(2015){Reffert}, {Bergmann}, {Quirrenbach}, {Trifonov}, \& {K{\"u}nstler}}]{ref15}
{Reffert}, S., {Bergmann}, C., {Quirrenbach}, A., {Trifonov}, T., \& {K{\"u}nstler}, A. 2015, \aap, 574, A116

\bibitem[{{Reiners} {et~al.}(2018){Reiners}, {Zechmeister}, {Caballero}, {Ribas}, {Morales}, {Jeffers}, {Sch{\"o}fer}, {Tal-Or}, {Quirrenbach}, {Amado}, {Kaminski}, {Seifert}, {Abril}, {Aceituno}, {Alonso-Floriano}, {Ammler-von Eiff}, {Antona}, {Anglada-Escud{\'e}}, {Anwand-Heerwart}, {Arroyo-Torres}, {Azzaro}, {Baroch}, {Barrado}, {Bauer}, {Becerril}, {B{\'e}jar}, {Ben{\'\i}tez}, {Berdinas̃}, {Bergond}, {Bl{\"u}mcke}, {Brinkm{\"o}ller}, {del Burgo}, {Cano}, {C{\'a}rdenas V{\'a}zquez}, {Casal}, {Cifuentes}, {Claret}, {Colom{\'e}}, {Cort{\'e}s-Contreras}, {Czesla}, {D{\'\i}ez-Alonso}, {Dreizler}, {Feiz}, {Fern{\'a}ndez}, {Ferro}, {Fuhrmeister}, {Galad{\'\i}-Enr{\'\i}quez}, {Garcia-Piquer}, {Garc{\'\i}a Vargas}, {Gesa}, {G{\'o}mez Galera}, {Gonz{\'a}lez Hern{\'a}ndez}, {Gonz{\'a}lez-Peinado}, {Gr{\"o}zinger}, {Grohnert}, {Gu{\`a}rdia}, {Guenther}, {Guijarro}, {de Guindos}, {Guti{\'e}rrez-Soto}, {Hagen}, {Hatzes}, {Hauschildt}, {Hedrosa}, {Helmling}, {Henning}, {Hermelo}, {Hern{\'a}ndez Arab{\'\i}},
  {Hern{\'a}ndez Casta{\~n}o}, {Hern{\'a}ndez Hernando}, {Herrero}, {Huber}, {Huke}, {Johnson}, {de Juan}, {Kim}, {Klein}, {Kl{\"u}ter}, {Klutsch}, {K{\"u}rster}, {Lafarga}, {Lamert}, {Lamp{\'o}n}, {Lara}, {Laun}, {Lemke}, {Lenzen}, {Launhardt}, {L{\'o}pez del Fresno}, {L{\'o}pez-Gonz{\'a}lez}, {L{\'o}pez-Puertas}, {L{\'o}pez Salas}, {L{\'o}pez-Santiago}, {Luque}, {Mag{\'a}n Madinabeitia}, {Mall}, {Mancini}, {Mandel}, {Marfil}, {Mar{\'\i}n Molina}, {Maroto Fern{\'a}ndez}, {Mart{\'\i}n}, {Mart{\'\i}n-Ruiz}, {Marvin}, {Mathar}, {Mirabet}, {Montes}, {Moreno-Raya}, {Moya}, {Mundt}, {Nagel}, {Naranjo}, {Nortmann}, {Nowak}, {Ofir}, {Oreiro}, {Pall{\'e}}, {Panduro}, {Pascual}, {Passegger}, {Pavlov}, {Pedraz}, {P{\'e}rez-Calpena}, {P{\'e}rez Medialdea}, {Perger}, {Perryman}, {Pluto}, {Rabaza}, {Ram{\'o}n}, {Rebolo}, {Redondo}, {Reffert}, {Reinhart}, {Rhode}, {Rix}, {Rodler}, {Rodr{\'\i}guez}, {Rodr{\'\i}guez-L{\'o}pez}, {Rodr{\'\i}guez Trinidad}, {Rohloff}, {Rosich}, {Sadegi}, {S{\'a}nchez-Blanco}, {S{\'a}nchez
  Carrasco}, {S{\'a}nchez-L{\'o}pez}, {Sanz-Forcada}, {Sarkis}, {Sarmiento}, {Sch{\"a}fer}, {Schmitt}, {Schiller}, {Schweitzer}, {Solano}, {Stahl}, {Strachan}, {St{\"u}rmer}, {Su{\'a}rez}, {Tabernero}, {Tala}, {Trifonov}, {Tulloch}, {Ulbrich}, {Veredas}, {Vico Linares}, {Vilardell}, {Wagner}, {Winkler}, {Wolthoff}, {Xu}, {Yan}, \& {Zapatero Osorio}}]{rei18}
{Reiners}, A., {Zechmeister}, M., {Caballero}, J.~A., {et~al.} 2018, \aap, 612, A49

\bibitem[{{Reyl{\'e}} {et~al.}(2021){Reyl{\'e}}, {Jardine}, {Fouqu{\'e}}, {Caballero}, {Smart}, \& {Sozzetti}}]{rey21}
{Reyl{\'e}}, C., {Jardine}, K., {Fouqu{\'e}}, P., {et~al.} 2021, \aap, 650, A201

\bibitem[{{Ribas} {et~al.}(2023){Ribas}, {Reiners}, {Zechmeister}, {Caballero}, {Morales}, {Sabotta}, {Baroch}, {Amado}, {Quirrenbach}, {Abril}, {Aceituno}, {Anglada-Escud{\'e}}, {Azzaro}, {Barrado}, {B{\'e}jar}, {Ben{\'\i}tez de Haro}, {Bergond}, {Bluhm}, {Calvo Ortega}, {Cardona Guill{\'e}n}, {Chaturvedi}, {Cifuentes}, {Colom{\'e}}, {Cont}, {Cort{\'e}s-Contreras}, {Czesla}, {D{\'\i}ez-Alonso}, {Dreizler}, {Duque-Arribas}, {Espinoza}, {Fern{\'a}ndez}, {Fuhrmeister}, {Galad{\'\i}-Enr{\'\i}quez}, {Garc{\'\i}a-L{\'o}pez}, {Gonz{\'a}lez-{\'A}lvarez}, {Gonz{\'a}lez Hern{\'a}ndez}, {Guenther}, {de Guindos}, {Hatzes}, {Henning}, {Herrero}, {Hintz}, {Huelmo}, {Jeffers}, {Johnson}, {de Juan}, {Kaminski}, {Kemmer}, {Khaimova}, {Khalafinejad}, {Kossakowski}, {K{\"u}rster}, {Labarga}, {Lafarga}, {Lalitha}, {Lamp{\'o}n}, {Lillo-Box}, {Lodieu}, {L{\'o}pez Gonz{\'a}lez}, {L{\'o}pez-Puertas}, {Luque}, {Mag{\'a}n}, {Mancini}, {Marfil}, {Mart{\'\i}n}, {Mart{\'\i}n-Ruiz}, {Molaverdikhani}, {Montes}, {Nagel}, {Nortmann},
  {Nowak}, {Pall{\'e}}, {Passegger}, {Pavlov}, {Pedraz}, {Perdelwitz}, {Perger}, {Ram{\'o}n-Ballesta}, {Reffert}, {Revilla}, {Rodr{\'\i}guez}, {Rodr{\'\i}guez-L{\'o}pez}, {Sadegi}, {S{\'a}nchez Carrasco}, {S{\'a}nchez-L{\'o}pez}, {Sanz-Forcada}, {Sch{\"a}fer}, {Schlecker}, {Schmitt}, {Sch{\"o}fer}, {Schweitzer}, {Seifert}, {Shan}, {Skrzypinski}, {Solano}, {Stahl}, {Stangret}, {Stock}, {St{\"u}rmer}, {Tabernero}, {Tal-Or}, {Trifonov}, {Vanaverbeke}, {Yan}, \& {Zapatero Osorio}}]{rib23}
{Ribas}, I., {Reiners}, A., {Zechmeister}, M., {et~al.} 2023, \aap, 670, A139

\bibitem[{Rousseeuw \& Croux(1993)}]{madfactor}
Rousseeuw, P.~J. \& Croux, C. 1993, Journal of the American Statistical Association, 88, 1273

\bibitem[{{Ryabchikova} {et~al.}(2015){Ryabchikova}, {Piskunov}, {Kurucz}, {Stempels}, {Heiter}, {Pakhomov}, \& {Barklem}}]{rad15}
{Ryabchikova}, T., {Piskunov}, N., {Kurucz}, R.~L., {et~al.} 2015, \physscr, 90, 054005

\bibitem[{{Sabotta} {et~al.}(2021){Sabotta}, {Schlecker}, {Chaturvedi}, {Guenther}, {Mu{\~n}oz Rodr{\'\i}guez}, {Mu{\~n}oz S{\'a}nchez}, {Caballero}, {Shan}, {Reffert}, {Ribas}, {Reiners}, {Hatzes}, {Amado}, {Klahr}, {Morales}, {Quirrenbach}, {Henning}, {Dreizler}, {Pall{\'e}}, {Perger}, {Azzaro}, {Jeffers}, {Kaminski}, {K{\"u}rster}, {Lafarga}, {Montes}, {Passegger}, \& {Zechmeister}}]{sab21}
{Sabotta}, S., {Schlecker}, M., {Chaturvedi}, P., {et~al.} 2021, \aap, 653, A114

\bibitem[{{Santos} {et~al.}(2017){Santos}, {Adibekyan}, {Dorn}, {Mordasini}, {Noack}, {Barros}, {Delgado-Mena}, {Demangeon}, {Faria}, {Israelian}, \& {Sousa}}]{san17}
{Santos}, N.~C., {Adibekyan}, V., {Dorn}, C., {et~al.} 2017, \aap, 608, A94

\bibitem[{{Santos} {et~al.}(2015){Santos}, {Adibekyan}, {Mordasini}, {Benz}, {Delgado-Mena}, {Dorn}, {Buchhave}, {Figueira}, {Mortier}, {Pepe}, {Santerne}, {Sousa}, \& {Udry}}]{san15}
{Santos}, N.~C., {Adibekyan}, V., {Mordasini}, C., {et~al.} 2015, \aap, 580, L13

\bibitem[{{Santos} {et~al.}(2001){Santos}, {Israelian}, \& {Mayor}}]{san01}
{Santos}, N.~C., {Israelian}, G., \& {Mayor}, M. 2001, \aap, 373, 1019

\bibitem[{{Schweitzer} {et~al.}(2019){Schweitzer}, {Passegger}, {Cifuentes}, {B{\'e}jar}, {Cort{\'e}s-Contreras}, {Caballero}, {del Burgo}, {Czesla}, {K{\"u}rster}, {Montes}, {Zapatero Osorio}, {Ribas}, {Reiners}, {Quirrenbach}, {Amado}, {Aceituno}, {Anglada-Escud{\'e}}, {Bauer}, {Dreizler}, {Jeffers}, {Guenther}, {Henning}, {Kaminski}, {Lafarga}, {Marfil}, {Morales}, {Schmitt}, {Seifert}, {Solano}, {Tabernero}, \& {Zechmeister}}]{sch19}
{Schweitzer}, A., {Passegger}, V.~M., {Cifuentes}, C., {et~al.} 2019, \aap, 625, A68

\bibitem[{{Seager} {et~al.}(2007){Seager}, {Kuchner}, {Hier-Majumder}, \& {Militzer}}]{sea07}
{Seager}, S., {Kuchner}, M., {Hier-Majumder}, C.~A., \& {Militzer}, B. 2007, \apj, 669, 1279

\bibitem[{{Shan} {et~al.}(2021){Shan}, {Reiners}, {Fabbian}, {Marfil}, {Montes}, {Tabernero}, {Ribas}, {Caballero}, {Quirrenbach}, {Amado}, {Aceituno}, {B{\'e}jar}, {Cort{\'e}s-Contreras}, {Dreizler}, {Hatzes}, {Henning}, {Jeffers}, {Kaminski}, {K{\"u}rster}, {Lafarga}, {Morales}, {Nagel}, {Pall{\'e}}, {Passegger}, {Rodriguez-L{\'o}pez}, {Schweitzer}, \& {Zechmeister}}]{sha21}
{Shan}, Y., {Reiners}, A., {Fabbian}, D., {et~al.} 2021, \aap, 654, A118

\bibitem[{{Smette} {et~al.}(2015){Smette}, {Sana}, {Noll}, {Horst}, {Kausch}, {Kimeswenger}, {Barden}, {Szyszka}, {Jones}, {Gallenne}, {Vinther}, {Ballester}, \& {Taylor}}]{sme15}
{Smette}, A., {Sana}, H., {Noll}, S., {et~al.} 2015, \aap, 576, A77

\bibitem[{{Sousa} {et~al.}(2011){Sousa}, {Santos}, {Israelian}, {Lovis}, {Mayor}, {Silva}, \& {Udry}}]{sou11}
{Sousa}, S.~G., {Santos}, N.~C., {Israelian}, G., {et~al.} 2011, \aap, 526, A99

\bibitem[{{Souto} {et~al.}(2017){Souto}, {Cunha}, {Garc{\'\i}a-Hern{\'a}ndez}, {Zamora}, {Allende Prieto}, {Smith}, {Mahadevan}, {Blake}, {Johnson}, {J{\"o}nsson}, {Pinsonneault}, {Holtzman}, {Majewski}, {Shetrone}, {Teske}, {Nidever}, {Schiavon}, {Sobeck}, {Garc{\'\i}a P{\'e}rez}, {G{\'o}mez Maqueo Chew}, \& {Stassun}}]{sou17}
{Souto}, D., {Cunha}, K., {Garc{\'\i}a-Hern{\'a}ndez}, D.~A., {et~al.} 2017, \apj, 835, 239

\bibitem[{{Souto} {et~al.}(2022){Souto}, {Cunha}, {Smith}, {Allende Prieto}, {Covey}, {Garc{\'\i}a-Hern{\'a}ndez}, {Holtzman}, {J{\"o}nsson}, {Mahadevan}, {Majewski}, {Masseron}, {Pinsonneault}, {Schneider}, {Shetrone}, {Stassun}, {Terrien}, {Zamora}, {Stringfellow}, {Lane}, {Nitschelm}, \& {Rojas-Ayala}}]{sou22}
{Souto}, D., {Cunha}, K., {Smith}, V.~V., {et~al.} 2022, \apj, 927, 123

\bibitem[{{Souto} {et~al.}(2018){Souto}, {Unterborn}, {Smith}, {Cunha}, {Teske}, {Covey}, {Rojas-Ayala}, {Garc{\'\i}a-Hern{\'a}ndez}, {Stassun}, {Zamora}, {Masseron}, {Johnson}, {Majewski}, {J{\"o}nsson}, {Gilhool}, {Blake}, \& {Santana}}]{sou18}
{Souto}, D., {Unterborn}, C.~T., {Smith}, V.~V., {et~al.} 2018, \apjl, 860, L15

\bibitem[{{Spina} {et~al.}(2021){Spina}, {Sharma}, {Mel{\'e}ndez}, {Bedell}, {Casey}, {Carlos}, {Franciosini}, \& {Vallenari}}]{spi21}
{Spina}, L., {Sharma}, P., {Mel{\'e}ndez}, J., {et~al.} 2021, Nature Astronomy, 5, 1163

\bibitem[{{Su{\'a}rez-Andr{\'e}s} {et~al.}(2018){Su{\'a}rez-Andr{\'e}s}, {Israelian}, {Gonz{\'a}lez Hern{\'a}ndez}, {Adibekyan}, {Delgado Mena}, {Santos}, \& {Sousa}}]{sua18}
{Su{\'a}rez-Andr{\'e}s}, L., {Israelian}, G., {Gonz{\'a}lez Hern{\'a}ndez}, J.~I., {et~al.} 2018, \aap, 614, A84

\bibitem[{{Tabernero} {et~al.}(2022){Tabernero}, {Marfil}, {Montes}, \& {Gonz{\'a}lez Hern{\'a}ndez}}]{tab22}
{Tabernero}, H.~M., {Marfil}, E., {Montes}, D., \& {Gonz{\'a}lez Hern{\'a}ndez}, J.~I. 2022, \aap, 657, A66

\bibitem[{{Talon} \& {Charbonnel}(2005)}]{tal05}
{Talon}, S. \& {Charbonnel}, C. 2005, \aap, 440, 981

\bibitem[{{Tsuji} {et~al.}(1996){Tsuji}, {Ohnaka}, \& {Aoki}}]{tsu96}
{Tsuji}, T., {Ohnaka}, K., \& {Aoki}, W. 1996, \aap, 305, L1

\bibitem[{{Virtanen} {et~al.}(2020){Virtanen}, {Gommers}, {Oliphant}, {Haberland}, {Reddy}, {Cournapeau}, {Burovski}, {Peterson}, {Weckesser}, {Bright}, {van der Walt}, {Brett}, {Wilson}, {Jarrod Millman}, {Mayorov}, {Nelson}, {Jones}, {Kern}, {Larson}, {Carey}, {Polat}, {Feng}, {Moore}, {Vand erPlas}, {Laxalde}, {Perktold}, {Cimrman}, {Henriksen}, {Quintero}, {Harris}, {Archibald}, {Ribeiro}, {Pedregosa}, {van Mulbregt}, \& {Contributors}}]{scipy}
{Virtanen}, P., {Gommers}, R., {Oliphant}, T.~E., {et~al.} 2020, Nature Methods, 17, 261

\bibitem[{{Weinberg} {et~al.}(2019){Weinberg}, {Holtzman}, {Hasselquist}, {Bird}, {Johnson}, {Shetrone}, {Sobeck}, {Allende Prieto}, {Bizyaev}, {Carrera}, {Cohen}, {Cunha}, {Ebelke}, {Fernandez-Trincado}, {Garc{\'\i}a-Hern{\'a}ndez}, {Hayes}, {J{\"o}nsson}, {Lane}, {Majewski}, {Malanushenko}, {M{\'e}sz{\'a}ros}, {Nidever}, {Nitschelm}, {Pan}, {Rix}, {Rybizki}, {Schiavon}, {Schneider}, {Wilson}, \& {Zamora}}]{wei19}
{Weinberg}, D.~H., {Holtzman}, J.~A., {Hasselquist}, S., {et~al.} 2019, \apj, 874, 102

\bibitem[{{Winters} {et~al.}(2015){Winters}, {Henry}, {Lurie}, {Hambly}, {Jao}, {Bartlett}, {Boyd}, {Dieterich}, {Finch}, {Hosey}, {Ianna}, {Riedel}, {Slatten}, \& {Subasavage}}]{win15}
{Winters}, J.~G., {Henry}, T.~J., {Lurie}, J.~C., {et~al.} 2015, \aj, 149, 5

\bibitem[{{Zechmeister} {et~al.}(2018){Zechmeister}, {Reiners}, {Amado}, {Azzaro}, {Bauer}, {B{\'e}jar}, {Caballero}, {Guenther}, {Hagen}, {Jeffers}, {Kaminski}, {K{\"u}rster}, {Launhardt}, {Montes}, {Morales}, {Quirrenbach}, {Reffert}, {Ribas}, {Seifert}, {Tal-Or}, \& {Wolthoff}}]{Zech18}
{Zechmeister}, M., {Reiners}, A., {Amado}, P.~J., {et~al.} 2018, \aap, 609, A12

\end{thebibliography}
\begin{appendix}
\section{Extra material}
\label{appendix}
{\onecolumn
\scriptsize
\begin{longtable}{lccccccccccc}
\caption{\label{tab:sample} 
Magnesium and silicon abundances and main parameters of investigated stars$^a$ (sorted by SpT). \tablefoot{$^a$Karmn identifier, spectral types (SpT), signal-to-noise ratio (S/N), projected rotational velocities ($\varv \sin{i}$), effective temperatures ($T_{\rm eff}$), surface gravities ($\log{g}$), metallicity ([Fe/H]), magnesium and silicon abundances ([Mg/H], [Si/H]), galactic population kinematic membership, and planet occurrence (N$_{\rm planets}$) for the stars analysed in this work. The full version of this table is available at the CDS.}} \\
\hline\hline\noalign{\smallskip}
Karmn & Name &  SpT & S/N & $\varv \sin{i}$ &  $T_{\rm eff}$ & $\log{g}$ & [Fe/H]   &[Mg/H] &  [Si/H]  & Pop. & N$_{\rm planets}$\\
      &      &      &     &  [km~s$^{-1}$]  &   [K]          &   [dex]   &  [dex]   & [dex] &  [dex]   &      &   \\
\noalign{\smallskip}\hline\noalign{\smallskip}
\endfirsthead
\caption{continued.}\\
\hline\hline\noalign{\smallskip}
Karmn & Name & SpT & S/N &$\varv \sin{i}$ &  $T_{\rm eff}$ & $\log{g}$ & [Fe/H] & [Mg/H] & [Si/H]  & Pop. & N$_{\rm planets}$ \\
      &      &     &     &  [km~s$^{-1}$] &   [K]          &   [dex]   &  [dex] & [dex]  &  [dex]   &     & \\
\noalign{\smallskip}\hline\noalign{\smallskip}
\endhead
\noalign{\smallskip}
\hline
\endfoot
J04167$-$120   & LP~714-47                 & K7.0\,V  & 258  & $\leq2.0$     & $3961\pm13$  & $5.01\pm0.09$  & $ 0.15\pm 0.04$  & $-0.02\pm 0.14$  & $ 0.22\pm 0.03$  & D    &  1 \\
J11110$+$304E  & HD~97101A                 & K7.0\,V  & 422  & $\leq2.0$     & $4211\pm13$  & $4.98\pm0.07$  & $ 0.04\pm 0.03$  & $ 0.03\pm 0.06$  & $ 0.08\pm 0.02$  & D    &  2 \\
J18198$-$019   & HD~168442                 & K7.0\,V  & 1447 & $\leq3.0$     & $4155\pm14$  & $5.01\pm0.06$  & $-0.14\pm 0.04$  & $-0.18\pm 0.04$  & $-0.14\pm 0.02$  & YD   &  0 \\
$\cdots$  & $\cdots$                  & $\cdots$  & $\cdots$ & $\cdots$   & $\cdots$  & $\cdots$  & $\cdots$ & $\cdots$ & $\cdots$  & $\cdots$  &  $\cdots$ \\
\end{longtable}}
\end{appendix}

\end{document}